\documentclass[useAMS,usenatbib]{mn2e}
\usepackage{graphicx}
\usepackage{amssymb}

\title[The ISM in the XMP H\,{\sc ii}/BCD galaxy Tol 65]
{On the properties of the interstellar medium in extremely metal-poor blue compact dwarf  galaxies. 
A VIMOS-IFU study of the cometary galaxy and Ly$\alpha$ absorber Tol 65.\thanks{Based on observations made with ESO Telescopes at the La Silla Paranal Observatory under programme 090.B-0242}}
\author[Lagos et al.]{P. Lagos$^{1}$\thanks{E-mail: plagos@astro.up.pt}, R. Demarco$^{2}$, P. Papaderos$^{1}$, E. Telles$^{3}$, A. Nigoche-Netro$^{4}$, 
\newauthor A. Humphrey$^{1}$, N. Roche$^{1}$ and J. M. Gomes$^{1}$\\
$^{1}$Instituto de Astrof\'isica e Ci\^encias do Espa\c{c}o, Universidade do Porto, CAUP, Rua das Estrelas, 4150-762 Porto, Portugal\\
$^{2}$Department of Astronomy, Universidad de Concepci\'on, Casilla 160-C, Concepci\'on, Chile\\
$^{3}$Observat\'{o}rio Nacional, Rua Jos\'{e} Cristino, 77, Rio de Janeiro, 20921-400, Brazil\\
$^{4}$Instituto de Astronom\'{\i}a y Meteorolog\'{\i}a, Av, Vallarta 2602. Col. Arcos Vallarta. Guadalajara, Jalisco. C.P. 44130, M\'exico}
\begin{document}

\date{Accepted 1988 December 15. Received 1988 December 14; in original form 1988 October 11}

\pagerange{\pageref{firstpage}--\pageref{lastpage}} \pubyear{2002}

\maketitle

\label{firstpage}

\begin{abstract}
In this study we present high-resolution VIsible Multi-Object Spectrograph integral field unit spectroscopy (VIMOS-IFU)
of the extremely metal-poor H\,{\sc ii}/blue compact dwarf (BCD) galaxy Tol 65. The optical appearance of this galaxy 
shows clearly a cometary morphology with a bright main body and an extended and diffuse stellar tail. 
We focus on the detection of metallicity gradients or inhomogeneities as expected if the ongoing star-formation activity 
is sustained by the infall/accretion of metal-poor gas. No evidences of significant spatial variations of abundances 
were found within our uncertainties. However, our findings show a slight anticorrelation between gas metallicity 
and star-formation rate at spaxel scales, in the sense that high star-formation is found in regions of low-metallicity, 
but the scatter in this relation indicates that the metals are almost fully diluted. 
Our observations show the presence of extended H$\alpha$ emission in the stellar tail of the galaxy. 
We estimated that the mass of the ionized gas in the tail M(H\,{\sc ii})$_{tail} \sim$1.7$\times$10$^5$ M$_{\odot}$ 
corresponds with $\sim$24 per cent of the total mass of the ionized gas in the galaxy.
We found that the H$\alpha$ velocity dispersion of the main body and the tail of the galaxy are comparable
with the one found in the neutral gas by previous studies. 
This suggests that the ionized gas still retains the kinematic memory of its parental cloud and likely a common origin.
Finally, we suggest that the infall/accretion of cold gas from the outskirts of the galaxy and/or 
minor merger/interaction may have produced the almost flat abundance gradient and the cometary morphology in Tol 65. 

\end{abstract}


\section{Introduction}

It is well established that interactions and/or galaxy collisions represent an important stage 
in the evolution of galaxies \citep[e.g.][]{Toomre1977,Sanders1988,Springel2000,Duc2011}. Numerous studies show clear indications 
of the importance of those external mechanisms for the enhancement of star-formation and 
its effects in the chemical enrichment of galaxies. In particular, in major merger of massive galaxies 
the preexisting gas metallicity can be substantially diluted by the inflow of metal poor gas 
from the outskirts to the nucleus \cite[e.g.][]{Rupke2010,Rich2012}.
In the case of low-mass, low-metallicity \citep[7.0 $\leq$ 12 + log(O/H) $\leq$ 8.4;][]{KunthSargent1983} 
and star-forming dwarf galaxies the effects of tidal interactions and/or mergers \cite[e.g.][]{Bekki2008,Verbeke2014}
also has a huge impact in their evolution. Observational evidences suggest that H\,{\sc ii}/BCD 
galaxies arise from the interactions or accretion of extended H\,{\sc i} cloud complexes 
\citep[e.g.][]{Taylor1993,Pustilnik2001,Ekta2008,EktaChengalur2010a}.
However, the triggering mechanisms of the current burst of star-formation in those objects 
is not yet clear since most are, apparently, isolated systems \citep[e.g.][]{TellesTerlevich1995,TellesMaddox2000}. 
Thus, if not triggered by external agents star-formation is likely produced by internal processes 
(e.g. gravitational cloud collapse, infall of gas in conjunction with small perturbations)
and/or minor mergers \citep[see][and references therein]{Lagos2011}.

As described above, a considerable fraction of these galaxies has been associated with H\,{\sc i} clouds 
\citep[][]{Taylor1993} or low-mass and undetected companions in the optical \citep[e.g.][]{Ekta2008}, 
which could rule out the idea of BCD galaxies as isolated systems \citep{Noeske2001}. 
In fact, a significant fraction of BCDs do show signs of extensions or tails in their outer envelopes, 
suggesting a tidal origin. Many of these low-metallicity galaxies that show ``cometary" or elongated shapes 
show values of 12+log(O/H) $<$ 7.6 \citep[e.g.][]{Papaderos2008}. 
Within this subsample of BCDs or extremely metal poor (XMP) BCD galaxies  
we found the least chemically evolved galaxies in the local Universe \citep{KunthOstlin2000}.
This particular morphology has been interpreted for high redshift galaxies in the Hubble Deep Field 
as the result of weak tidal interactions \cite[][]{Straughn2006},  
gravitational instabilities in gas-rich and turbulent galactic disks in formation at high redshift \cite[][]{Bournaud2009} 
and stream-driven accretion of metal-poor gas from the cosmic web \cite[][]{DekelBirnboim2006,Dekel2009}.
\cite{Papaderos2008} argue that weak interactions between low-mass stellar or gaseous companions, 
or propagating shock waves, lead to a bar-like gas distribution triggering the star-formation that by propagation 
could subsequently produce a cometary morphology in XMP BCDs.
 
Recently, \cite{Sanchez2013,Sanchez2014a} interpret the metallicity variation in a sample of low metallicity 
galaxies with cometary morphology as a sign of external gas accretion/infall of metal poor gas.
They argue that these results are consistent with the local ``tadpole" galaxies being disks in early
stages of assembling, with their star-formation sustained by pristine gas infall.
In any case, dwarf galaxies tend to show flat abundance (O/H, N/O) gradients 
\citep[e.g.][]{KobulnickySkillman1997,KobulnickySkillman1998,LeeSkillman2004,Lagos2009,Lagos2012},
suggesting efficient dispersion and mixing of metals in the interstellar medium (ISM)
by expanding starburst-driven superbubbles \citep[e.g.][]{TenorioTagle1996}, afterward the gas 
begins to cool down by radiation and gravity, and/or external gas infall \citep[e.g.][]{PerezMontero2011,Lagos2012}. 
These mechanisms have been put forth as potential 
causes for the observed flat metal distributions in local dwarf galaxies. 
While in massive star-forming and/or interacting galaxies, bar-induced rotation 
or shear \citep[e.g.][]{RoyKunth1995} and merger-induced gas flows \citep[e.g.][]{Rupke2010} 
could produce the metal dispersal and mixing.
As expressed above, local H\,{\sc ii}/BCD and XMP BCD galaxies are considered chemically homogeneous 
and only in a few isolated cases we observed indications of variation of 12+log(O/H) over the ISM 
\citep[e.g. SBS 0335-052E, Haro 11, HS 2236+1344;][]{Izotov1997,Izotov2006,Guseva2012,Lagos2014}. 

In addition to the expected low metal content in young galaxies at high redshift, according to theoretical models
these objects should produce strong Ly$\alpha$ (1216 $\rm \AA$) emission as the result of their intense star-formation 
activity \citep[e.g.][]{PartridgePeebles1967,CharlotFall1993}. 
However, the absence and/or diminished Ly$\alpha$ emission in these galaxies, which is
significantly lower than the theoretical recombination ratio, indicate that
the Ly$\alpha$ photons are likely redistributed by multiple scattering in the H\,{\sc i} envelope, or are absorbed by dust. 
Examples of the detection of Ly$\alpha$ halos produced 
by H\,{\sc i} scattering envelopes can be seen in the literature \citep[e.g.][]{Humphrey2013,Guaita2015}.  
It has been suggested in the literature \citep[e.g.][]{Atek2008} that there should be an increase in the
Ly$\alpha$/H$\beta$ flux ratio as the metallicity of the galaxy decreases, 
since presumably low-metallicity objects contain less dust and hence suffer less Ly$\alpha$ photon destruction. 
Ly$\alpha$ can also be enhanced at low gas metallicity due to collisional excitation \citep{VillarMartin2007}.
As pointed by \cite{Izotov2004}, the fact that there is no Ly$\alpha$ emission in the two most 
metal-deficient BCDs known, \,{\sc i} Zw 18 \citep{Kunth1994} and SBS 0335-052 \citep{Thuan1997}, and also in Tol 65, 
argues against the existence of some correlation between the Ly$\alpha$ emission, metallicity and dust 
\citep[][]{Giavalisco1996,Lequeux1995}. 
Therefore, an important issue in the understanding of Ly$\alpha$ emission in galaxies is the study 
of the spatial distribution of properties in the ISM of those objects in order to see the different 
regulation mechanisms involved in the detectability of Ly$\alpha$ emission.

Although during the last years some progress has been made in this field, many questions remain 
about the metal content, the mechanism involved in the transport and mixing of metals 
and the star-formation activity in H\,{\sc ii}/BCD and XMP galaxies. 
The morphologically diverse nature of H\,{\sc ii}/BCD, and XMP BCDs, galaxies allows us 
to consider the role played by galaxy interactions and the feedback between 
the star-formation and the ISM in the observed metal distributions.
Our main objective in this paper is to carry out a spatial
investigation of the warm gas properties in the ``cometary" and Ly$\alpha$ absorbing
XMP H\,{\sc ii}/BCD galaxy Tol 65. These include, the spatial distribution of emission lines, 
equivalent width EW(H$\beta$), extinction c(H$\beta$), ionization ratios ([O\,{\sc iii}]$\lambda$5007/H$\beta$, 
[S\,{\sc ii}]$\lambda\lambda$6717,6731/H$\alpha$ and [N\,{\sc ii}]$\lambda$6584/H$\alpha$), 
kinematics, the chemical pattern (e.g. O/H, N/H and N/O, etc.) 
and also the possible dependence between these properties. 
Our aim is to search for the existence of metallicity inhomogeneities
as expected if the ongoing star-formation activity is sustained by the infall or accretion of metal-poor gas. 
To this end, we use high resolution integral field unit (IFU) spectroscopy observations.

The paper is organized as follows: Sect. \ref{sect_tol65} describes the most important properties
of our analyzed galaxy. Sect. \ref{sect_obser_reduc} contains the
technical details regarding observations and the data reduction. 
Sect. \ref{sect_results} describes our results: the ionized gas structure as well as the physical 
and kinematic properties of the ionized gas. Sect. \ref{sect_discussion} discusses the results. 
Finally, Sect. \ref{conclusions} itemizes the conclusions.

\section[]{Overview of Tol 65}\label{sect_tol65}

\begin{figure*}
\includegraphics[width=180mm]{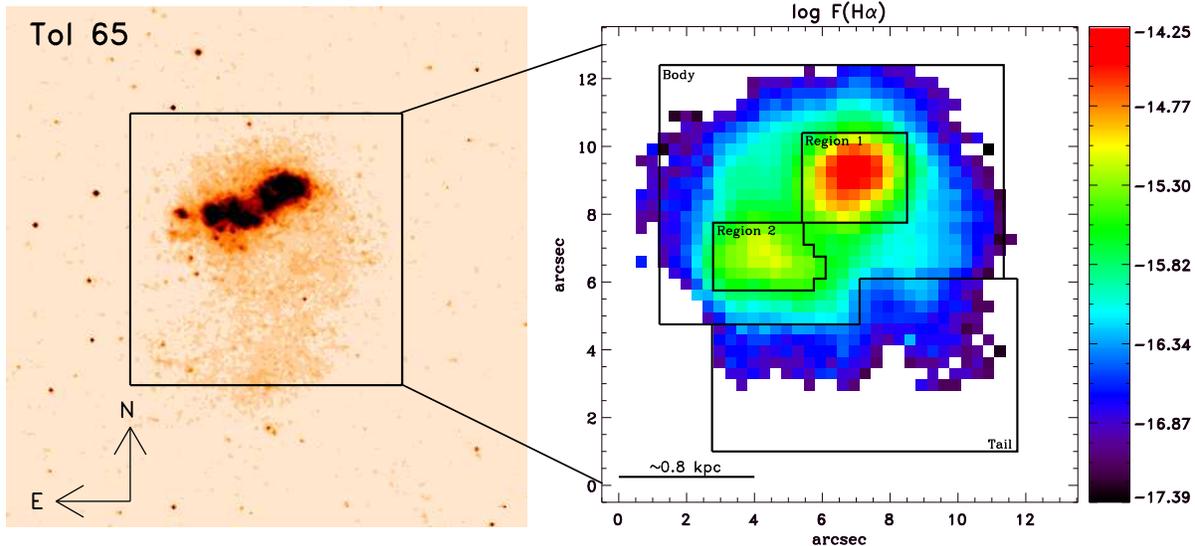}
  \caption{HST image of Tol 65 in the filter F435W (left-hand panel). 
H$\alpha$ emission line map, in log, of the galaxy (right-hand panel). 
The VIMOS-IFU FoV of 13\arcsec$\times$13$\arcsec (\sim$2691 pc $\times$ 2691 pc) is shown over the images.
Fluxes are in units of ergs cm$^{-2}$ s$^{-1}$.
}
\label{figure_images_ha}
\end{figure*}

\begin{table*}
 \centering
 \begin{minipage}{130mm}
  \caption{General parameters of Tol 65.}
  \label{table_parameters}
  \begin{tabular}{@{}lccccccrlr@{}}
  \hline
Parameter                      & Value                         &Reference \\       
\hline 
Other names                    & Tol 1223-359                  &\\
                               &  ESO380-G027                  &\\
RA  (J2000)                    & 12$^{h}$25$^{m}$46.9$^{s}$    & \\                      
DEC (J2000)                    & -36$^{o}$14$^{m}$1$^{s}$      &\\
Distance (Mpc)                 &  42.7                         &Obtained from NED\\
Pixel scale (pc/arcsec)        &  207                          &Obtained from NED\\
z                              &  0.00936008                   &Derived from the present observations\\
c(H$\beta$)                    &  0.29                         &Derived from the present observations\\
12 + log(O/H)                  &  7.56$\pm$0.06                &Derived from the present observations\\
log M$_{HI}$ (M$_{\odot}$)     & 8.82                          &\cite{Smoker2000}, \cite{PustilnikMartin2007}\\
N(HI) (cm$^{-2}$)              & (2.5$\pm$1.0)$\times$10$^{21}$&\cite{ThuanIzotov1997} \\
\hline
\end{tabular}
\end{minipage}
\end{table*}

Tol 65 is one of the most metal-deficient BCD galaxies known. 
\cite{Izotov2004} derived a low oxygen abundance of 12+log(O/H)=7.54$\pm$0.01, 
while from the present observations we have obtained a value of 12+log(O/H)=7.56$\pm$0.06.
The Nitrogen-to-oxygen ratio log(N/O)=-1.60$\pm$0.02 obtained by Izotov et al. appears to be in agreement 
with the observed value in other XMP BCD galaxies.
The optical appearance of this galaxy (see left-hand panel in Fig. \ref{figure_images_ha}) 
shows a clear cometary shape with a bright main body and an extended and diffuse stellar tail 
\cite[see also Fig. 4 in][]{Lagos2007}.
Therefore, Tol 65 is classified as a iI,C according to the \cite{LooseThuan1986} classification scheme.  
Narrow band images \citep[e.g. H$\beta$;][]{Lagos2007} show that the main body of the galaxy is 
composed of two Giant H\,{\sc ii} regions (GH\,{\sc ii}Rs), or star cluster complexes, 
while Very Large Telescope (VLT) \citep{Papaderos1999} and  Hubble Space Telescope (HST) images of the same galaxy 
(left-hand panel of Fig. \ref{figure_images_ha}) show that these GH\,{\sc ii}Rs are formed by several star clusters. 
The age of the underlying stellar component inferred from radially averaged colour profiles, 
assuming an instantaneous burst, by \cite{Papaderos1999} and \cite{Noeske2003} is of the order of 10$^8$ yr. 
\cite{ThuanIzotov1997} showed that the spectrum of this galaxy shows a broad absorption-damped 
Ly$\alpha$ profile similar to the most metal poor galaxies known 
in the local Universe: \,{\sc i} Zw18 \citep[][]{Kunth1994} and SBS 0335-052 \citep[][]{ThuanIzotov1997}.
\cite{Atek2008} argue that the young age of Tol 65 ($\sim$3 Myr) is consistent with 
a young starburst embedded in a static H\,{\sc i} cloud, which produces a damped absorption.
Narrow band Ly$\alpha$ imaging of this galaxy, from the same study, show a diffuse Ly$\alpha$ morphology 
consistent with strong absorption/scattering of Ly$\alpha$ photons by H\,{\sc i}. 
The basic characteristics of the galaxy are compiled in Table \ref{table_parameters}.

\section[]{Observations and data reduction}\label{sect_obser_reduc}

The data have been obtained in service mode with the IFU mode of the VIsible Multi-Object Spectrograph (VIMOS), 
hereafter VIMOS-IFU, at ESO mounted on the 8.2 m VLT UT3/Melipal telescope at the Paranal Observatory 
in Chile. The VIMOS-IFU array is composed by 4 quadrants and covered by 1600 fibres. 
We used the scale on the sky of 0.33$\arcsec$ per fiber covering a Field of View (FoV) 
of 13\arcsec $\times$ 13\arcsec. We used the high-resolution blue (HR-blue; 0.71 $\rm\AA$ pixel$^{-1}$) 
and orange (HR-orange; 0.62 $\rm\AA$ pixel$^{-1}$) spectral setups, offering a spectral 
resolving power R=1440 between $\sim$3755--5346$\rm \AA$ and R=2650 between 
$\sim$5244--7440 $\rm\AA$, respectively. One Arc-line and three flat-field calibration frames were taken 
for every observing block (OB). The observing log can be found in Table \ref{table_observing_log}.
The data reduction was carried out using the \textit{esorex} software version 3.10.2. 
This includes bias subtraction, flat-field correction, wavelength calibration and flux calibration. 
The flux calibration was done using observations of spectrophotometric standards included 
in the standard VIMOS calibration plan. 
Sky subtraction can be a problem with the VIMOS-IFU, as there are no sky-dedicated fibres. 
We obtained a night sky spectra from a third exposure, within each set of OB, 
dithered by +0.00$\arcsec$ in RA and by +0.48$\arcsec$ in DEC. 
Therefore, the sky subtraction was performed by averaging the spectra recorded by the sky fibres and subtracting this 
spectrum from that of each spaxel in the VIMOS-IFU exposures.
The data cubes obtained using the gratings HR-orange and HR-blue were shifted and combined, 
using a sigma-clipping algorithm to remove the cosmic rays, forming a final data cube
covering a total spectral range from $\sim$3755 to $\sim$7480 $\rm \AA$. 
We correct for the quadrant-to-quadrant intensity differences following the procedure applied by 
\cite{Lagerholm2012} assuming that the intensity correction is uniform within each quadrant. 
Therefore, we re-normalized the quadrants by comparing the intensity levels of the neighbouring 
pixels at the quadrant borders and taking a first-order gradient in intensity across the borders into account. 
When comparing the mean intensity value in quadrants Q1, Q3, and Q4 with respect to quadrant Q2 we found
a value of $\sim$0.2, 0.7 and 0.3, respectively.
We checked the effects of the atmospheric refraction (DAR) in each one of our data cubes
using an IRAF-based script \citep{WalshRoy1990}, but there is no evidence for significant DAR in our data.

\begin{figure*}
\includegraphics[width=88mm]{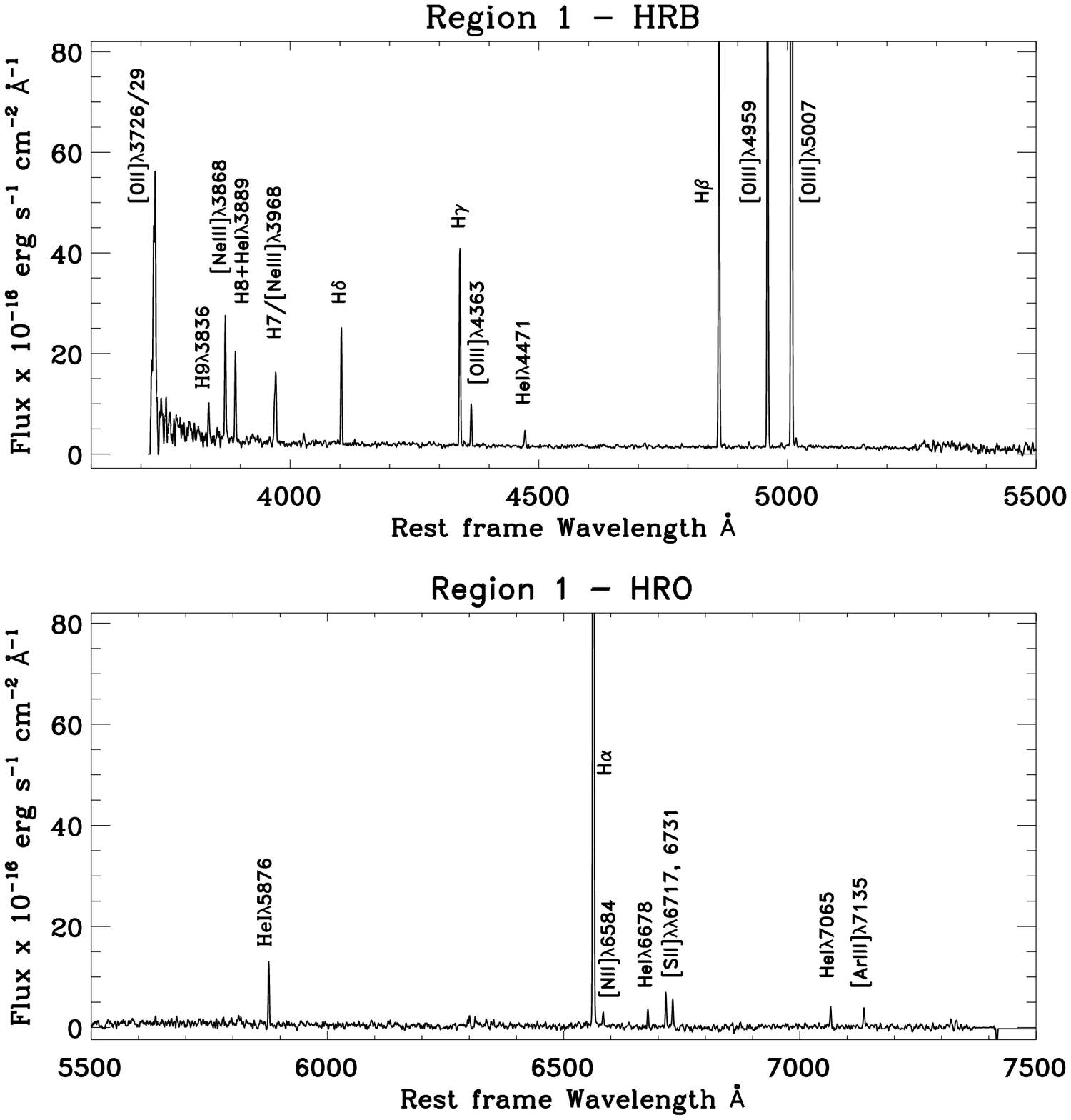}
\includegraphics[width=88mm]{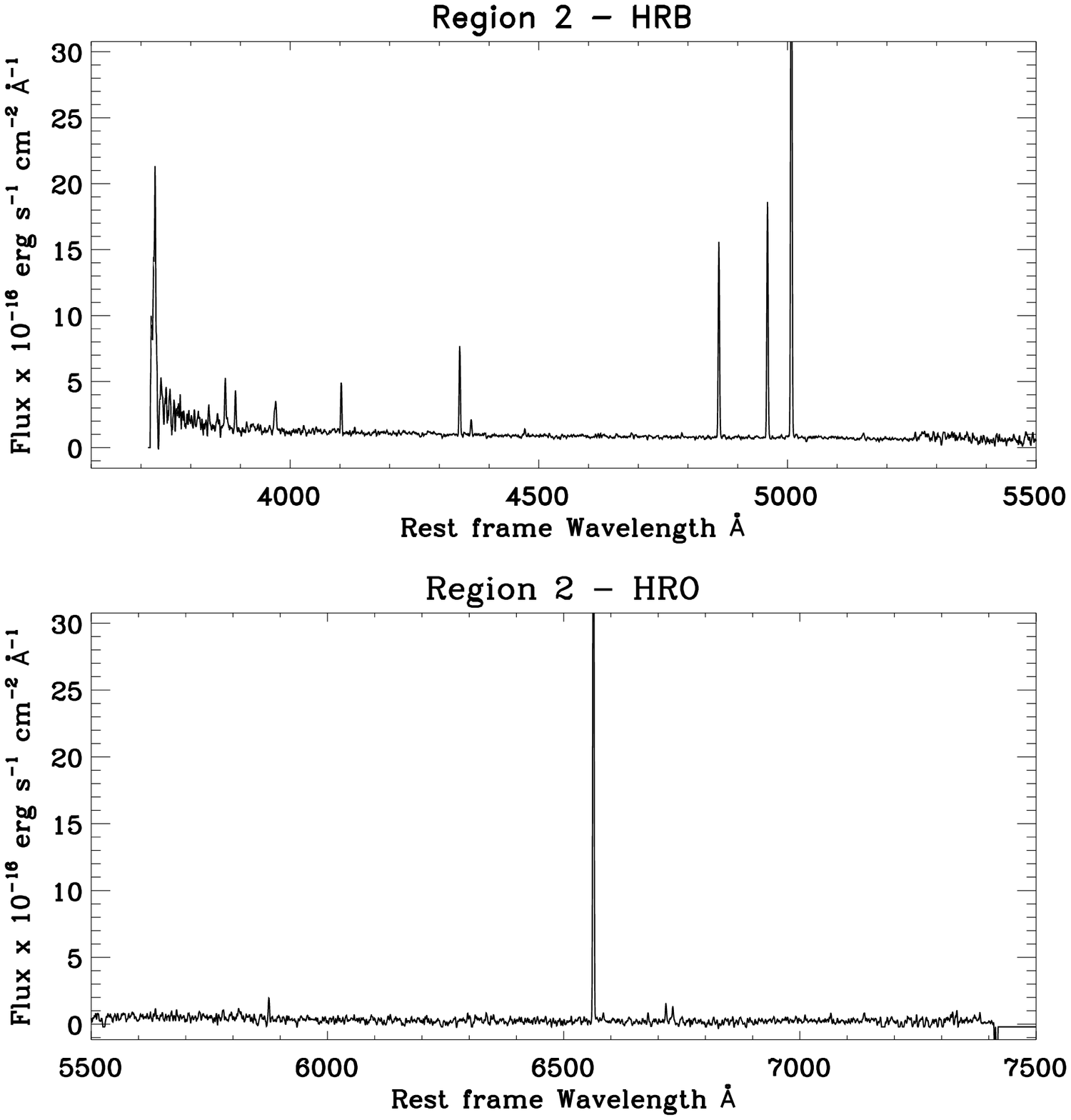}
 \caption{Tol 65. Integrated HR-blue (upper panel) and HR-orange (lower panel) spectrum of the two 
 GH\,{\sc ii}Rs resolved in the galaxy. The most important emission lines are labelled in the panels.}
\label{figure_spectra_Tol65}
\end{figure*}

Finally, the emission line fluxes were measured using the IRAF task
\textit{fitprofs} by fitting, spaxel by spaxel, Gaussian profiles. 
The two main star cluster complexes or GH\,{\sc ii}Rs, called here regions nos. 1 and 2,
and the other apertures considered in this study, the body and tail, are defined 
in the right-hand panel of Fig. \ref{figure_images_ha}. Figure \ref{figure_spectra_Tol65} shows the integrated spectra 
obtained summing up all spaxels of regions nos. 1 and 2, respectively.
In this figure we identified the main emission lines detected and used in our study.
In Table 3 we show the observed F($\lambda$) and corrected emission line fluxes I($\lambda$) 
relative to the H$\beta$ and their errors \citep[see][]{Lagos2009} multiplied by a factor of 100, for the main body 
of the galaxy and regions no. 1 and 2, respectively. 
We also indicate, in this table, the observed flux of the H$\beta$ emission line, 
the EW(H$\beta$), and the extinction coefficient c(H$\beta$) for the aforementioned apertures. 
The instrumental contribution to the line broadening was obtained by
fitting a single Gaussian to isolated arc lines on a wavelength calibrated arc exposure in the HR-orange observations. 
Thus, we found the resolution to be FWHM=1.91 $\rm \AA$ ($\sim$87.48 km s$^{-1}$) near H$\alpha$.

\begin{table}
 \centering
 \begin{minipage}{140mm}
  \caption{Observing log.}
  \label{table_observing_log}
  \begin{tabular}{@{}lcccccccccc@{}}
  \hline
 Grating &OB & Date      &Exp. time    & Airmass\footnote{Mean value at start of exposures and at end of exposures} & Seeing\footnote{Mean value during the observation} \\
         &   &           & (s)         &                                                     &                                                                \\
 \hline
HRO      & 1 & 2012-12-28&2$\times$932  & 1.70-1.57  & 0.63\\ 
         & 2 & 2013-12-28&              & 1.38-1.31  & 1.05\\ 
         & 3 & 2013-01-21&              & 1.06-1.04  & 1.03\\ 
HRB      & 1 & 2013-03-15&              & 1.35-1.43  & 1.56\\
         & 2 & 2013-04-04&              & 1.04-1.03  & 0.89\\
         & 3 & 2013-04-04&              & 1.02-1.02  & 0.89\\         
\hline
\end{tabular}
\end{minipage}
\end{table}

\section{Results}\label{sect_results}

\subsection[]{Emission lines, EW(H$\beta$), extinction and emission line ratio maps}\label{sect_I_c_EW}

The peak of H$\alpha$ emission is placed on the designated region no. 1 
while faint H$\alpha$ emission is detected over the stellar tail of the galaxy. 
This structure is visible more easily in the lower panel of Fig. \ref{figure_emission_tail_Tol65}.
We will present this Fig. below in Sect. \ref{sect_tail_gasmass}. 
The recombination lines (H$\beta$, H$\gamma$, etc.) and the forbidden emission lines [O\,{\sc iii}]$\lambda$5007, 
[O\,{\sc iii}]$\lambda$4363, [S\,{\sc ii}]$\lambda\lambda$6712,6731, [N\,{\sc ii}]$\lambda$6584, etc. 
displays very similar spatial morphology to that of H$\alpha$ in Fig. \ref{figure_images_ha}. 

\begin{figure*}
\includegraphics[width=85mm]{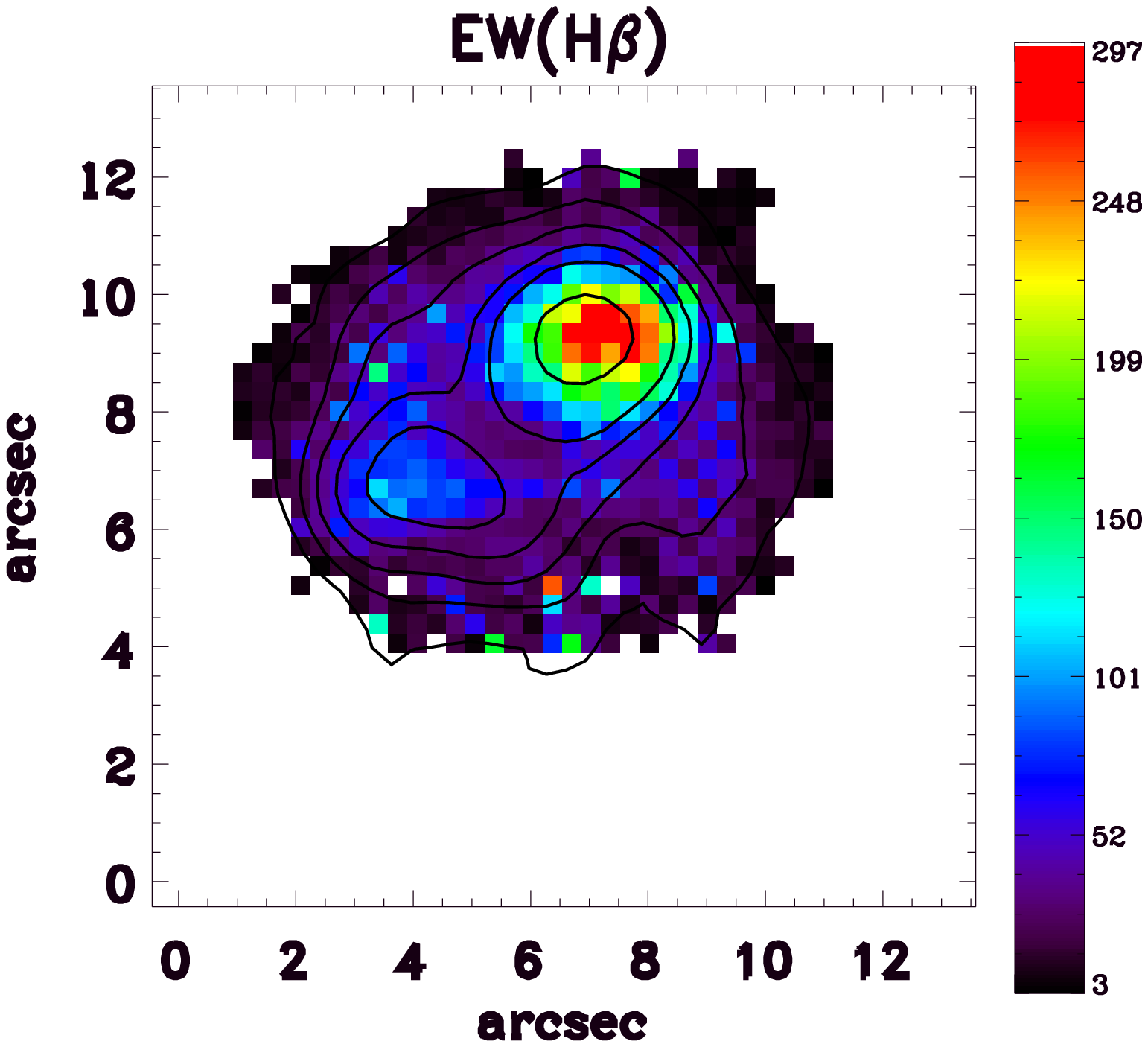}
\includegraphics[width=85mm]{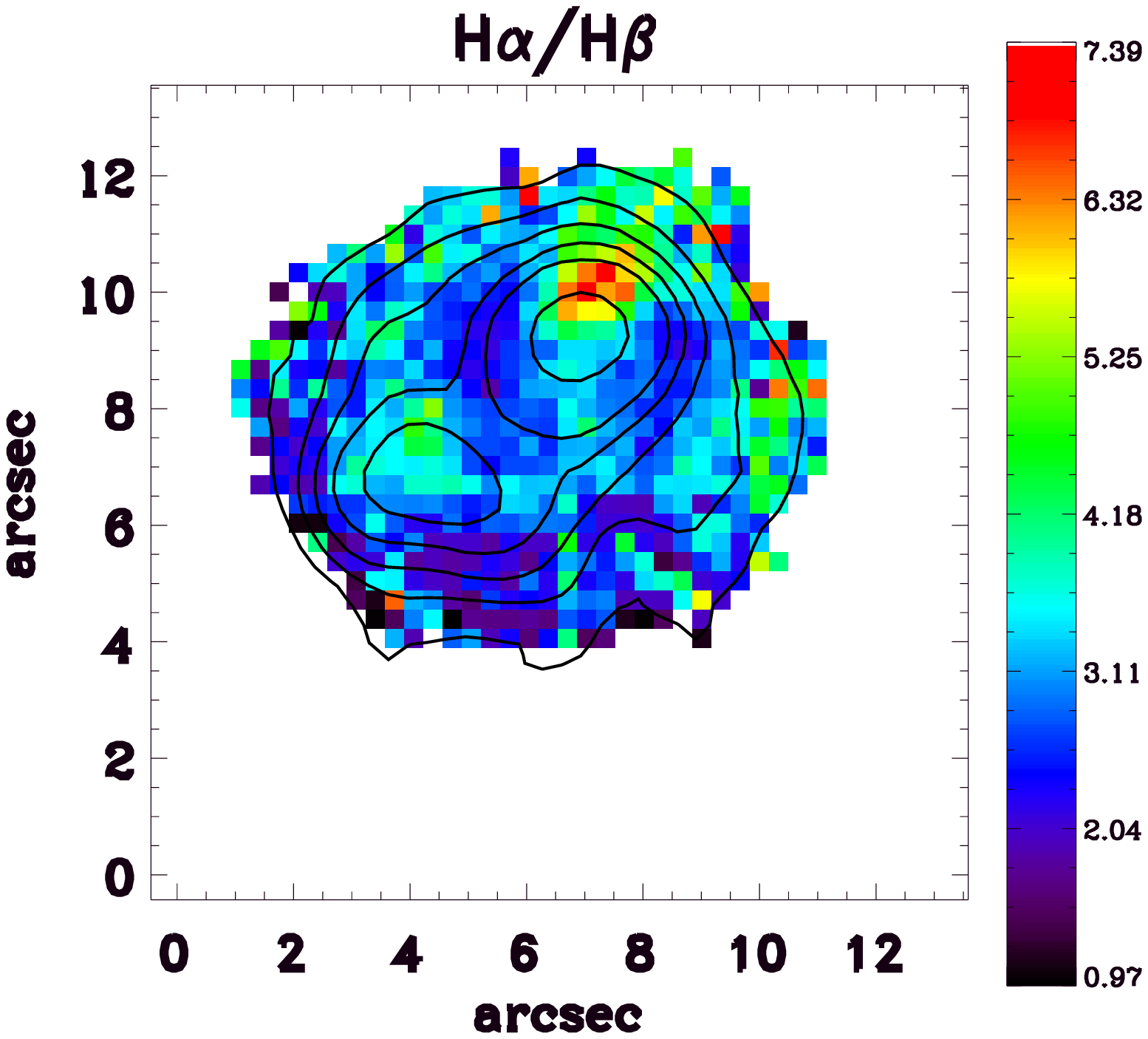}
  \caption{Right-hand panel: Spatial distribution of the H$\alpha$/H$\beta$ emission line ratio.
  Left-hand panel: H$\beta$ equivalent width map of the galaxy. EW(H$\beta$) is in units of $\rm \AA$.
  The H$\alpha$ emission line contours are overlaid in the panels. North is up and east is to the left.
  }
  \label{figure_ew_hahb}
\end{figure*}

The H$\beta$ equivalent width, EW(H$\beta$), map of the galaxy in Fig. \ref{figure_ew_hahb} (left-hand panel) 
ranges values from $\sim$3 $\rm \AA$ to 297 $\rm \AA$, with the highest values located at the two 
star-forming regions. This map resembles that obtained by \cite{Lagos2007}, it to say, the peak 
of the EW(H$\beta$) and H$\alpha$ emission of the galaxy are almost coincident. 
Although the observed EW(H$\beta$) can be affected by the contribution to the continuum 
by the accumulation of old populations of stars from previous bursts, multiplicity of the star-forming knots into 
several individual star clusters \citep[][]{Lagos2007}, and the leaking of ionizing photons 
\citep[see][and references therein]{Papaderos2013}, its value can be understood as an age indicator 
\citep[][]{Dottori1981,Copetti1986}, with EW(H$\beta$) values decreasing with the age of the starburst. 
Hence, a crude estimation of the age of the current burst of star-formation 
can be obtained by using the integrated EW(H$\beta$) in each of the regions.
Therefore, we use the same SSP models from STARBURST99 \citep{Leitherer1999} as in \cite{Lagos2014} 
in order to obtain a rough idea of the age of the current burst of star-formation in the GH\,{\sc ii}Rs of the galaxy.
From the integrated spectrum of the body of the galaxy one infers, for an instantaneous burst, an upper limit to the age 
of 4.26 Myr ($\sim$4 Myr) and  2.91 Myr ($\sim$3 Myr) and 4.95 Myr ($\sim$5 Myr) for regions nos. 1 and 2, respectively.
The aforementioned effects themselves represent an important source of uncertainty in the determination 
of absolute ages and also it may be highly model dependent. However, those results are indicative 
of a young burst and are only provided for the sake of comparison with previous studies \citep[e.g.][]{Atek2008}.

The reddening coefficient c(H$\beta$) was determined from the comparison of the observed 
flux ratio of the Balmer H$\alpha$ and H$\beta$ and the theoretical line ratio 
computed by \cite{OsterbrockFerland2006} for the physical conditions of the nebula assuming 
case B at T=2$\times$10$^{4}$ K.
We used the reddening function f($\lambda$) derived by \cite{Cardelli1989} and assuming R$_{V}$ = 3.1. 
Figure \ref{figure_ew_hahb} (right-hand panel) shows the spatial distribution of the H$\alpha$/H$\beta$ emission 
line ratio. The c(H$\beta$) value obtained in this study is 0.29$\pm$0.05 for the main body and 0.40$\pm$0.02 
and 0.29$\pm$0.06 for regions nos. 1 and 2, respectively. 
Our values are similar, within the uncertainties, to the ones obtained by \cite{Guseva2011} 
for their individual regions with values of 0.285$\pm$0.018 and 0.180$\pm$0.019, respectively. 
The c(H$\beta$) distribution as seen by the H$\alpha$/H$\beta$ map does not resemble 
the structure of the dust content in the extinction, E(B-V), map of Tol 65 presented by \cite{Atek2008}.
Our spatial distribution shows the maximum values displaced from the peak of H$\alpha$ line emission, 
while the one presented by \cite{Atek2008} appears similar to the H$\alpha$ emission line distribution.

\begin{figure*}
\includegraphics[width=85mm]{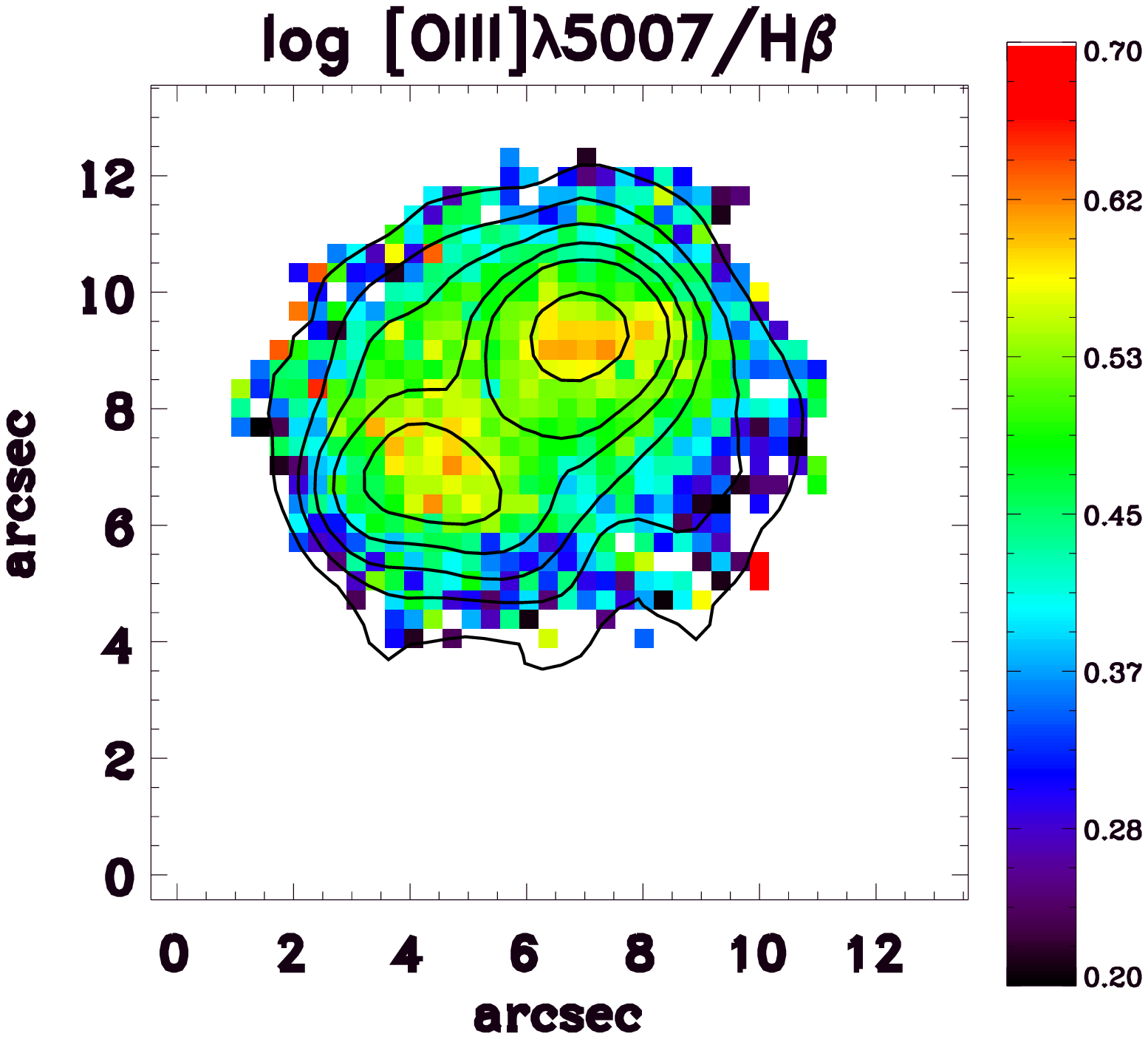}
\includegraphics[width=85mm]{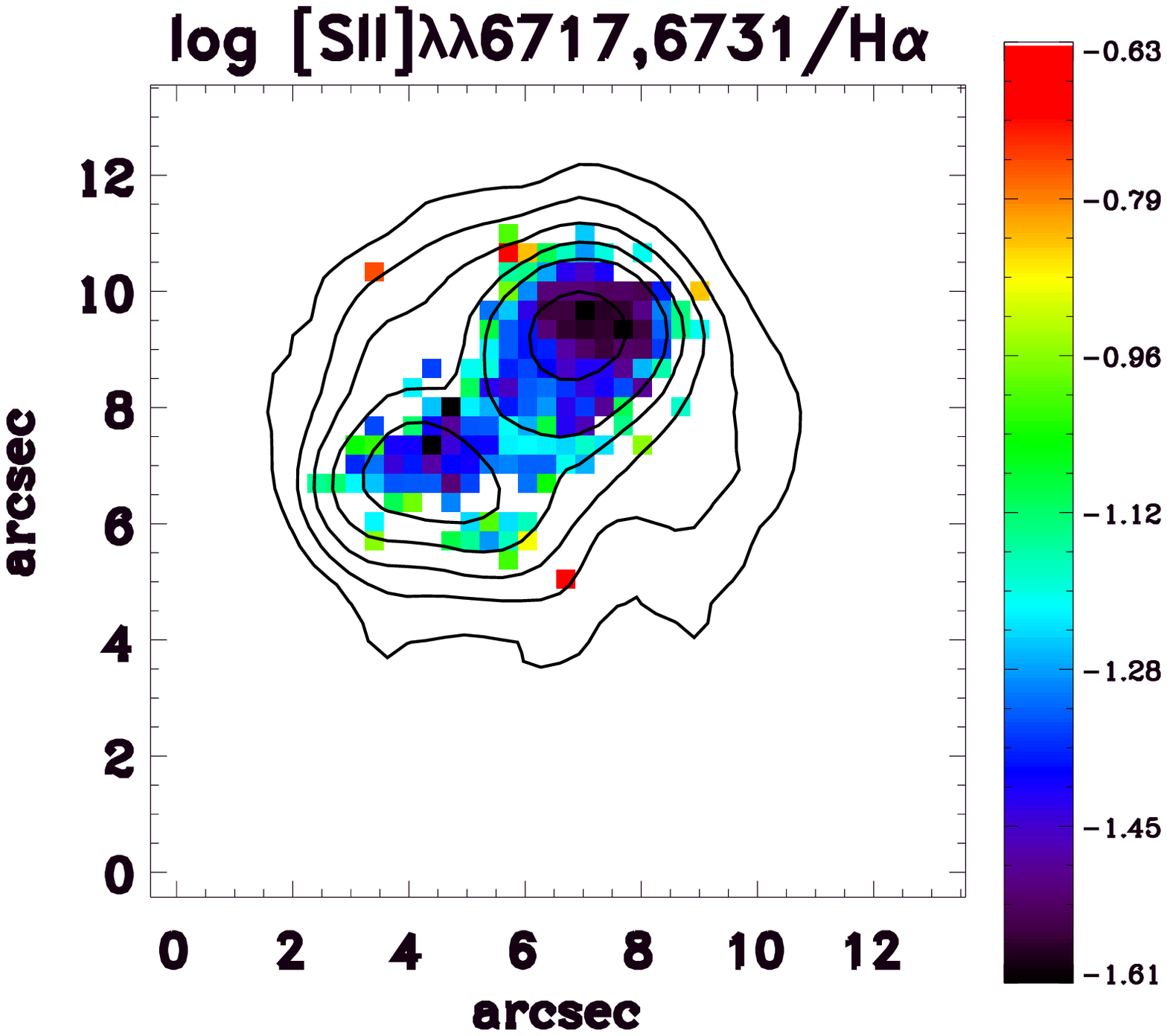}\\
\includegraphics[width=85mm]{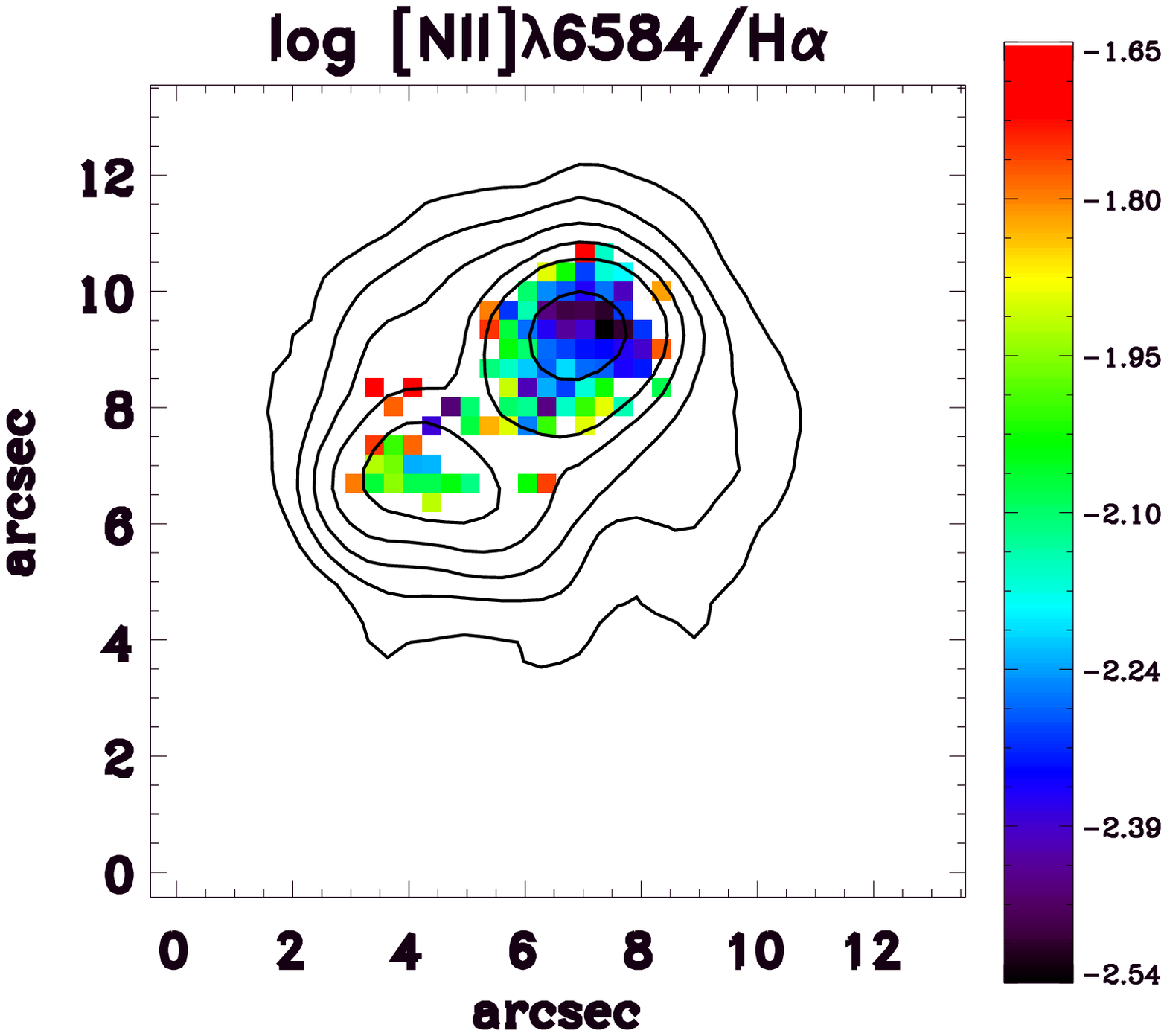}
  \caption{[O\,{\sc iii}]$\lambda$5007/H$\beta$, [S\,{\sc ii}]$\lambda\lambda$6712,6731/H$\alpha$ and  
  [N\,{\sc ii}]$\lambda$6584/H$\alpha$  emission line ratio maps. H$\alpha$ emission line contours are overlaid in the panels.
  North is up and east is to the left.}
  \label{figure_ratios_galaxies}
\end{figure*}

To infer the dominant ionization source in the gas at spaxel scales we employ the commonly used diagnostic 
or BPT \citep{Baldwin1981} diagrams on the basis of the following emission-line ratios: 
[O\,{\sc iii}]$\lambda$5007/H$\beta$, [S\,{\sc ii}]$\lambda\lambda$6717,6731/H$\alpha$ and 
[N\,{\sc ii}]$\lambda$6582/H$\alpha$. In Figure \ref{figure_ratios_galaxies} we show those emission-line-ratio maps.
The spatial structure of those maps changes from the peak of H$\alpha$ emission, in each one of the regions, to the outer 
part of the galaxy. Meanwhile the ionization structure on the GH\,{\sc ii}Rs is rather constant for all emission line ratios.
All points fall, in Fig. \ref{figure_ratios_galaxies}, in the locus predicted by models of photo-ionization 
(we will not show the BPT diagrams in this study) by young stars in H\,{\sc ii} regions \citep{OsterbrockFerland2006}  
corroborating that photoionization from stellar sources is the dominant excitation mechanism in Tol 65.

We found a weak He\,{\sc ii} $\lambda$4686 emission in both regions of the galaxy 
(we will not show the spectra in this study). It shows the same shapes as the spectra 
showed by \cite{Izotov2004} in their fig. 3 and 4, but the low S/N=signal-to-noise (S/N) ratio $<$ 3 
in these regions does not allow us an adequate 
study of the spatial morphology of this emission line. However, we note that the peak of 
H$\alpha$ and He \,{\sc ii} $\lambda$4686 are coincident, so we can associate the spatial
distribution of He \,{\sc ii} $\lambda$4686, in Tol 65, with current sites of star-formation activity.
Given the very young age of the starburst $\sim$3-5 Myr the presence of Wolf-Rayet (WR) stars is expected.
Since the population of WRs decreases with metallicity \citep{CervinoMasHesse1994} and the detection of WR features
depends on the quality of the spectra, location through the galaxy and size of the apertures, 
the stellar feature of WR stars are not observed in the spectra of Tol 65. In any case, the lack of detection of WR features 
in this galaxy does not exclude the presence of these stars.

Finally, as a check on the results we compare our spectrophotometric values with those obtained using
narrow-band H$\beta$ images by \cite{Lagos2007}. In \cite{Lagos2007} we found a F(H$\beta$) of 
6.91$\times$10$^{-14}$ erg s$^{-1}$cm$^{-2}$ for the main body of the galaxy 
and 3.47$\times$10$^{-14}$ erg s$^{-1}$cm$^{-2}$ and 0.41$\times$10$^{-14}$ erg s$^{-1}$cm$^{-2}$ 
for regions nos. 1 and 2, respectively. 
While the EW(H$\beta$), in that study, reaches values of 157 $\rm \AA$, 287 $\rm \AA$ and 86 $\rm \AA$ 
for the same apertures. Our results agree very well with the aforementioned study with an average 
difference of $\sim$30 per cent for H$\beta$ fluxes and $\sim$22 per cent for EW(H$\beta$). 
Our integrated F(H$\alpha$) flux, in the body of the galaxy, differ $\sim$35 per cent with that obtained by \cite{GildePaz2003} 
from Narrow-band H$\alpha$ images. These differences are perfectly explained by the uncertainties 
associated with measurements and the different sizes adopted for the apertures.  
However, the flux calibration accuracy has little effect on the analysis of the internal physical conditions 
of this one object.

  \begin{table*}
 \begin{minipage}{150mm}
  \caption{Observed and extinction corrected emission lines in Tol 65. 
The fluxes are relative to F(H$\beta$)=100.}
  \label{table_integrated_fluxes}
  \begin{tabular}{@{}lccccccccr@{}}
  \hline
 & \multicolumn{2}{c}{Main Body} & \multicolumn{2}{c}{Region no. 1} & \multicolumn{2}{c}{Region no. 2}  \\
 &  F($\lambda$)/F(H$\beta$)  & I($\lambda$)/I(H$\beta$)&  F($\lambda$)/F(H$\beta$)  & I($\lambda$)/I(H$\beta$) &  F($\lambda$)/F(H$\beta$)  & I($\lambda$)/I(H$\beta$)   \\
 \hline
$\left[OII\right] \lambda$3726    & 31.55$\pm$2.63 & 39.13$\pm$4.61 & 45.94$\pm$7.35 & 61.82$\pm$13.99 & 21.89$\pm$1.48 & 27.15$\pm$2.59\\
$\left[OII\right] \lambda$3729    & 50.48$\pm$6.34 & 62.58$\pm$11.11& 46.31$\pm$4.41 & 62.28$\pm$8.39  & 63.79$\pm$1.29 & 79.08$\pm$2.26\\
H9$\lambda$3836                   & 13.64$\pm$2.05 & 16.65$\pm$3.54 &  7.79$\pm$0.40 & 10.26$\pm$0.74  & 10.76$\pm$1.73 & 13.13$\pm$2.98\\
$\left[Ne III\right] \lambda$3868 & 15.65$\pm$3.92 & 19.01$\pm$6.73 & 23.97$\pm$2.94 & 31.35$\pm$5.44  & 19.09$\pm$4.53 & 23.19$\pm$7.78\\
H8+He I $\lambda$3889             & 16.87$\pm$3.32 & 20.42$\pm$5.68 & 17.87$\pm$1.05 & 23.26$\pm$1.93  & 16.52$\pm$4.46 & 20.00$\pm$7.64\\
$\left[Ne III\right] \lambda$3968 &  5.48$\pm$0.94 &  6.55$\pm$1.59 &  8.64$\pm$0.22 & 11.04$\pm$0.40  &  8.88$\pm$0.87 & 10.61$\pm$1.47\\
H7  $\lambda$3970                 & 17.49$\pm$5.58 & 20.89$\pm$9.42 & 14.77$\pm$3.42 & 18.87$\pm$6.18  & 21.26$\pm$5.23 & 25.39$\pm$8.83\\
H$\delta  \lambda$4101            & 26.48$\pm$2.04 & 30.87$\pm$3.36 & 25.54$\pm$1.10 & 31.56$\pm$1.92  & 26.29$\pm$2.14 & 30.65$\pm$3.68\\
H$\gamma  \lambda$4340            & 42.80$\pm$2.70 & 47.52$\pm$4.24 & 44.51$\pm$1.80 & 51.41$\pm$2.94  & 42.69$\pm$2.59 & 47.40$\pm$4.07\\
$\left[OIII\right] \lambda$4363   &  7.71$\pm$0.33 &  8.52$\pm$0.51 &  9.25$\pm$0.21 & 10.61$\pm$0.34  &  8.11$\pm$0.46 &  8.96$\pm$0.72\\
HeI $\lambda$4471                 &  2.76$\pm$0.95 &  2.98$\pm$1.45 &  3.46$\pm$0.66 &  3.85$\pm$1.02  &  2.82$\pm$0.88 &  3.05$\pm$1.34\\
H$\beta \lambda$4861              &100.00$\pm$0.70 &100.00$\pm$0.99 &100.00$\pm$0.81 &100.00$\pm$1.15 &100.00$\pm$3.37 & 100.00$\pm$4.77\\
$\left[OIII\right] \lambda$4959   &118.10$\pm$1.89 &116.09$\pm$2.63 &126.94$\pm$1.40 &123.97$\pm$1.93 &122.46$\pm$2.84 & 120.38$\pm$3.95 \\
$\left[OIII\right] \lambda$5007   &352.26$\pm$5.36 &343.49$\pm$7.39 &377.43$\pm$2.63 &364.53$\pm$3.59 &369.05$\pm$6.22 & 359.86$\pm$8.58 \\
HeI $\lambda$5876                 & 14.27$\pm$2.57 & 12.46$\pm$3.17 & 13.22$\pm$1.24 & 10.96$\pm$1.45 & 13.95$\pm$2.19 &  12.18$\pm$2.70 \\
$\left[OI\right] \lambda$6300     &  $\cdots$  & $\cdots$   &  2.08$\pm$0.23 &  1.63$\pm$0.25 &  $\cdots$ &$\cdots$\\
$\left[SIII\right] \lambda$6312   &  $\cdots$  & $\cdots$   &  1.83$\pm$0.34 &  1.43$\pm$0.38 &  $\cdots$ &$\cdots$\\
$\left[OI\right] \lambda$6363     &  $\cdots$  & $\cdots$   &  1.15$\pm$0.45 &  0.89$\pm$0.49 & $\cdots$     & $\cdots$\\
H$\alpha \lambda$6563             &348.56$\pm$6.00 &285.75$\pm$6.96 &377.43$\pm$2.63 &286.95$\pm$2.83 &348.63$\pm$6.71 &285.80$\pm$7.78\\
$\left[NII\right] \lambda$6584    &  3.52$\pm$0.80 &  2.88$\pm$0.93 &  3.29$\pm$0.41 &  2.49$\pm$0.44 &  3.66$\pm$0.55 &  2.99$\pm$0.64 \\
HeI $\lambda$6678                 & 11.45$\pm$1.60 &  9.29$\pm$1.84 &  4.83$\pm$0.37 &  3.62$\pm$0.39 &  7.05$\pm$0.82 &  5.72$\pm$0.94 \\
$\left[SII\right] \lambda$6717    &  7.16$\pm$0.83 &  5.79$\pm$0.95 &  6.81$\pm$0.39 &  5.08$\pm$0.41 &  7.61$\pm$0.96 &  6.15$\pm$1.10    \\
$\left[SII\right] \lambda$6731    &  5.81$\pm$0.67 &  4.69$\pm$0.76 &  5.52$\pm$0.13 &  4.11$\pm$0.24 &  4.71$\pm$0.75 &  3.80$\pm$0.86   \\
HeI $\lambda$7065                 &  5.46$\pm$0.80 &  4.28$\pm$0.89 &  4.44$\pm$0.58 &  3.17$\pm$0.59 &  3.25$\pm$0.95 &  2.55$\pm$1.05  \\
$\left[ArIII\right] \lambda$7136  &  3.12$\pm$0.55 &  2.43$\pm$0.61 &  3.91$\pm$0.33 &  2.77$\pm$0.33 &  3.79$\pm$0.59 &  2.95$\pm$0.65  \\
                                  &       \\
F(H$\beta$)\footnote{In units of $\times$10$^{-15}$ erg cm$^{-2}$ s$^{-1}$} &45.96$\pm$0.16   & &   27.21$\pm$0.11 & &  4.75$\pm$0.08  \\
EW(H$\beta$)\footnote{In units of $\rm \AA$}                                & 121& &257& & 68 \\ 
c(H$\beta$)                                                                 & 0.29$\pm$0.05 &  & 0.40$\pm$0.02 & & 0.29$\pm$0.06  & \\
\hline

\end{tabular}
\end{minipage}
\end{table*}

\subsection[]{Chemical abundances}

\subsubsection[]{Electron temperature and density}\label{sect_temp_den}

The first step in the abundance derivation is to obtain the electron density and temperature.
The [O\,{\sc iii}]$\lambda\lambda$4959,5007/[O\,{\sc iii}]$\lambda$4363 intensity ratio was used to
determine the electron temperature t$_e$(O\,{\sc iii}) and the electron density n$_e$(S\,{\sc ii}) was obtained   
from the emission line doublet [S\,{\sc ii}]$\lambda$6716/[S\,{\sc ii}]$\lambda$6731 ratio.
We computed the values of t$_e$(O\,{\sc iii}) and n$_e$(S\,{\sc ii}) using the IRAF STS package \textit{nebular}. 

\begin{figure*}
\includegraphics[width=85mm]{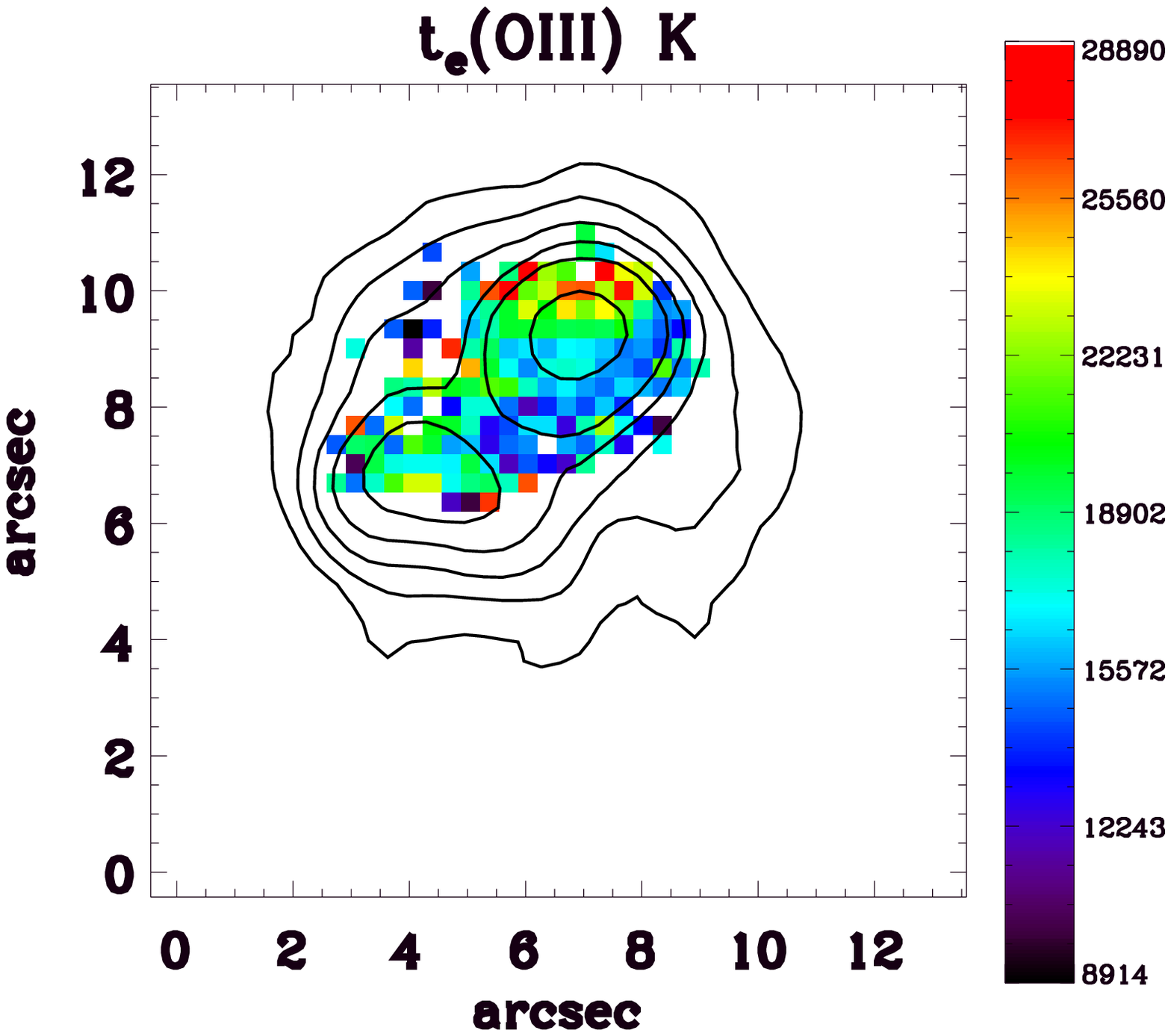}
\includegraphics[width=85mm]{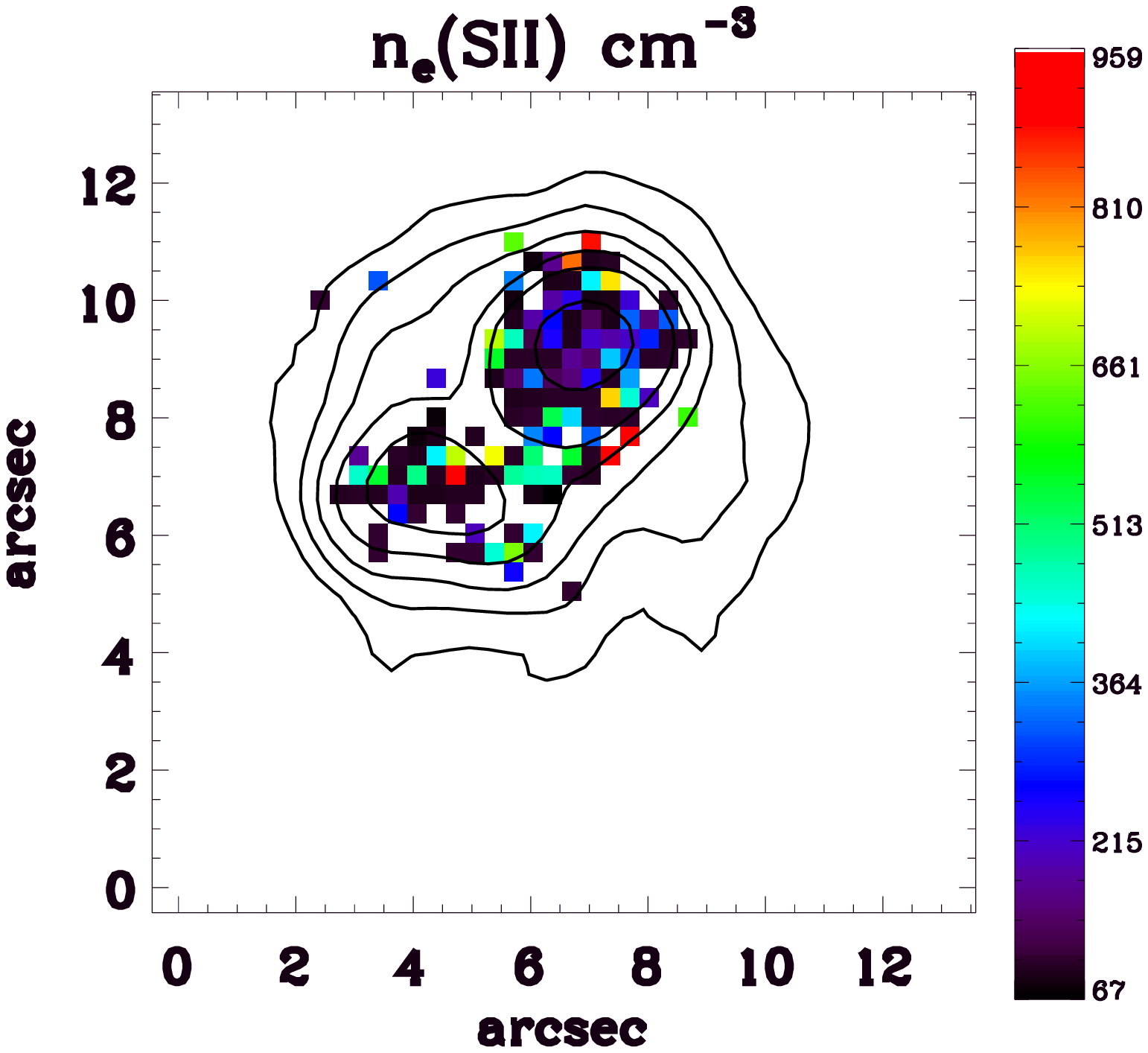}
  \caption{Electron temperature t$_e$(O\,{\sc iii}) (left-hand panel) and density n$_e$(S\,{\sc ii}) (right-hand panel) 
  maps of Tol 65. H$\alpha$ emission line contours are overlaid in the panel. North is up and east is to the left.}
  \label{figure_temp_den_galaxies}
\end{figure*}

Figure \ref{figure_temp_den_galaxies} shows the spatial distribution of electron temperature t$_e$(O\,{\sc iii}) 
(left-hand panel) and density n$_e$(S\,{\sc ii}) (right-hand panel) of Tol 65. 
The range of valid data points for the t$_e$(O\,{\sc iii}) varies 
from 0.89 to $\sim$2.9 $\times$ 10$^4$ K, while the electron density range from 
$\sim$67 cm$^{-3}$ to $\sim$1000 cm$^{-3}$. We note that
the ratio [S\,{\sc ii}]$\lambda$6717/[S\,{\sc ii}]$\lambda$6731 is typically greater than 1 
and only a few spaxels show values $\sim$1000 cm$^{-3}$, which indicates a dominant
low density regime \citep{OsterbrockFerland2006} in the ISM of the galaxy, hence 
we assumed an electron density of n$_{e}\sim$100 cm$^{-3}$ for those apertures in our calculations. 
Aperture regions nos. 1 and 2 show temperatures of 18587$\pm$335 and 17152$\pm$715 K, while n$_e$(S\,{\sc ii})  
shows values of 212 cm$^{-3}$ and 100 cm$^{-3}$, respectively. The t$_e$(O\,{\sc iii}) and n$_e$(S\,{\sc ii}) 
found in this study agree, within the uncertainties, with the ones obtained by \cite{Izotov2004} 
and \cite{Guseva2011} derived from their high and medium-resolution spectra. 

\subsubsection[]{Abundances determination}\label{sect_abundances}

Oxygen, nitrogen, neon, sulfur and argon abundances were calculated using 
the five level atomic model FIVEL \citep[][]{DeRobertis1987} implemented in the IRAF STS package \textit{nebular} 
\citep[\textit{abund};][]{ShawDufour1994} using the t$_e$(O\,{\sc iii}) and n$_e$(S\,{\sc ii}) 
obtained previously in Sect. \ref{sect_temp_den} and assuming that:
$\frac{O}{H}=\frac{O^{+}}{H^{+}} + \frac{O^{+2}}{H^{+}}$, $\frac{N}{H}=ICF(N) \frac{N^{+}}{H^{+}}$,
$\frac{Ne}{H}=ICF(Ne) \frac{Ne^{+2}}{H^{+}}$, $\frac{S}{H}=ICF(S) \left(\frac{S^{+}}{H^{+}}+\frac{S^{+2}}{H^{+}}\right)$,
and $\frac{Ar}{H}=ICF(Ar) \frac{Ar^{+2}}{H^{+}}$.
Where  O$^{+}$, O$^{+2}$, N$^{+}$, Ne$^{+2}$, S$^{+}$ and Ar$^{+2}$ correspond to the ions 
of the different atomic species, and the ICF to the ionization correction factor, respectively. 
We used the relationship between S$^{+}$ and S$^{+2}$ given by \cite{KingsburhBarlow1994}, assuming an ICF(Ar)=1.87 
and that the t$_{e}$(O \,{\sc ii}) temperature is given by 
t$_{e}$(O \,{\sc ii}) = 2/(t$_{e}^{-1}$(O \,{\sc iii})+0.8) \citep{Pagel1992}. 
In Figure \ref{figure_abundances_Tol65} we show the oxygen, nitrogen, neon, argon and sulfur abundance maps and
in Fig. \ref{figure_abundances_ratio_Tol65} we show the log(N/O), log(Ne/O), log(Ar/O) and log(S/O) ratio maps of Tol 65.
Table \ref{table_abundances_Tol65} shows the abundances calculated for the different apertures considered in this study.
Below in section \ref{sect_spatial_abund} we will describe the distribution of abundances for the different atomic species 
obtained from the procedures described above.

\begin{table*}
 \centering
 \begin{minipage}{85mm}
  \caption{Ionic, abundances and integrated properties of Tol 65.
   ICFs are taken from Kingsburgh \& Barlow (1994)
   and assuming ICF(Ar)=1.87.}
  \label{table_abundances_Tol65}
 \begin{tabular}{@{}lccc@{}}
  \hline
                                &  Main Body &   Region no. 1 & Region no. 2\\
 \hline
Te(OIII) K                      &17087$\pm$534 &18587$\pm$336&17152$\pm$715\\
Ne(SII) cm$^{-3}$               &100           &212          &  100        \\
O$^{+}$/H$^{+} \times$10$^{5}$  & 0.98$\pm$0.04 & 1.09$\pm$0.02 & 1.02$\pm$0.06\\
O$^{++}$/H$^{+} \times$10$^{5}$ & 2.67$\pm$0.19 & 2.35$\pm$0.09 & 2.76$\pm$0.26\\
O/H $\times$10$^{5}$            & 3.64$\pm$0.23 & 3.44$\pm$0.11 & 3.78$\pm$32\\
12+log(O/H)                     & 7.56$\pm$0.06 & 7.54$\pm$0.04 & 7.58$\pm$0.08\\
N$^{+}$/H$^{+} \times$10$^{6}$  & 0.24$\pm$0.01 & 0.20$\pm$0.01 & 0.25$\pm$0.01\\
ICF(N)                          & 3.73$\pm$0.40 & 3.15$\pm$0.18 & 3.72$\pm$0.52\\
N/H $\times$10$^{6}$            & 0.90$\pm$0.12 & 0.62$\pm$0.04 & 0.93$\pm$0.17\\
12+log(N/H)                     & 5.95$\pm$0.14 & 5.79$\pm$0.07 & 5.97$\pm$0.18\\
log(N/O)                        &-1.61$\pm$0.20 &-1.74$\pm$0.10 &-1.61$\pm$0.27\\
S$^{+}$/H$^{+} \times$10$^{7}$  & $\cdots$  & 0.59$\pm$0.03 & $\cdots$\\
S$^{++}$/H$^{+} \times$10$^{6}$ & $\cdots$  & 0.36$\pm$0.02 & $\cdots$\\
ICF(S)                          & 1.18$\pm$0.05 & 1.14$\pm$0.03 & 1.18$\pm$0.06\\
S/H  $\times$10$^{6}$           & $\cdots$  & 0.47$\pm$0.01 & $\cdots$\\
12+log(S/H)                     & $\cdots$  & 5.67$\pm$0.03 & $\cdots$\\
log(S/O)                        & $\cdots$  &-1.86$\pm$0.06 & $\cdots$\\
Ne$^{++}$/H$^{+} \times$10$^{5}$& 0.33$\pm$0.03 & 0.44$\pm$0.02 & 0.40$\pm$0.04\\
ICF(Ne)                         & 1.37$\pm$0.19 & 1.46$\pm$0.11 & 1.37$\pm$0.25\\
Ne/H $\times$10$^{5}$           & 0.46$\pm$0.10 & 0.65$\pm$0.08 & 0.55$\pm$0.17\\
12+log(Ne/H)                    & 6.66$\pm$0.23 & 6.81$\pm$0.12 & 6.74$\pm$0.31\\
log(Ne/O)                       &-0.90$\pm$0.29 &-0.72$\pm$0.15 &-0.83$\pm$0.40\\
Ar$^{++}$/H$^{+} \times$10$^{7}$& 0.78$\pm$0.04 & 0.79$\pm$0.02 & 0.95$\pm$0.06\\
Ar/H $\times$10$^{7}$           & 1.47$\pm$0.07 & 1.48$\pm$0.04 & 1.77$\pm$0.12\\
12+log(Ar/H)                    & 5.17$\pm$0.05 & 5.17$\pm$0.03 & 5.25$\pm$0.07\\
log(Ar/O)                       &-2.39$\pm$0.11 &-2.37$\pm$0.06 &-2.33$\pm$0.15\\
\hline
log($\left[O III\right]\lambda$5007/H$\beta$)            &   0.53$\pm$0.02 &  0.56$\pm$0.01& 0.56$\pm$0.02\\
log($\left[N II\right]\lambda$6584/H$\alpha$)            &  -1.54$\pm$0.34 & -1.60$\pm$0.18 & -1.52$\pm$0.21\\
log($\left[S II\right]\lambda\lambda$6717,6731/H$\alpha$)&  -0.98$\pm$0.16 & -1.04$\pm$0.07 & -1.00$\pm$0.19\\
$\sigma$(H$\alpha$) km s$^{-1}$                          &   20.8$\pm$0.6  &  19.6$\pm$0.6  &  22.5$\pm$0.4 \\
\hline
\end{tabular}
\end{minipage}
\end{table*}

\subsubsection{Elemental abundances and its spatial distribution}\label{sect_spatial_abund}

\begin{figure*}
\includegraphics[width=58mm]{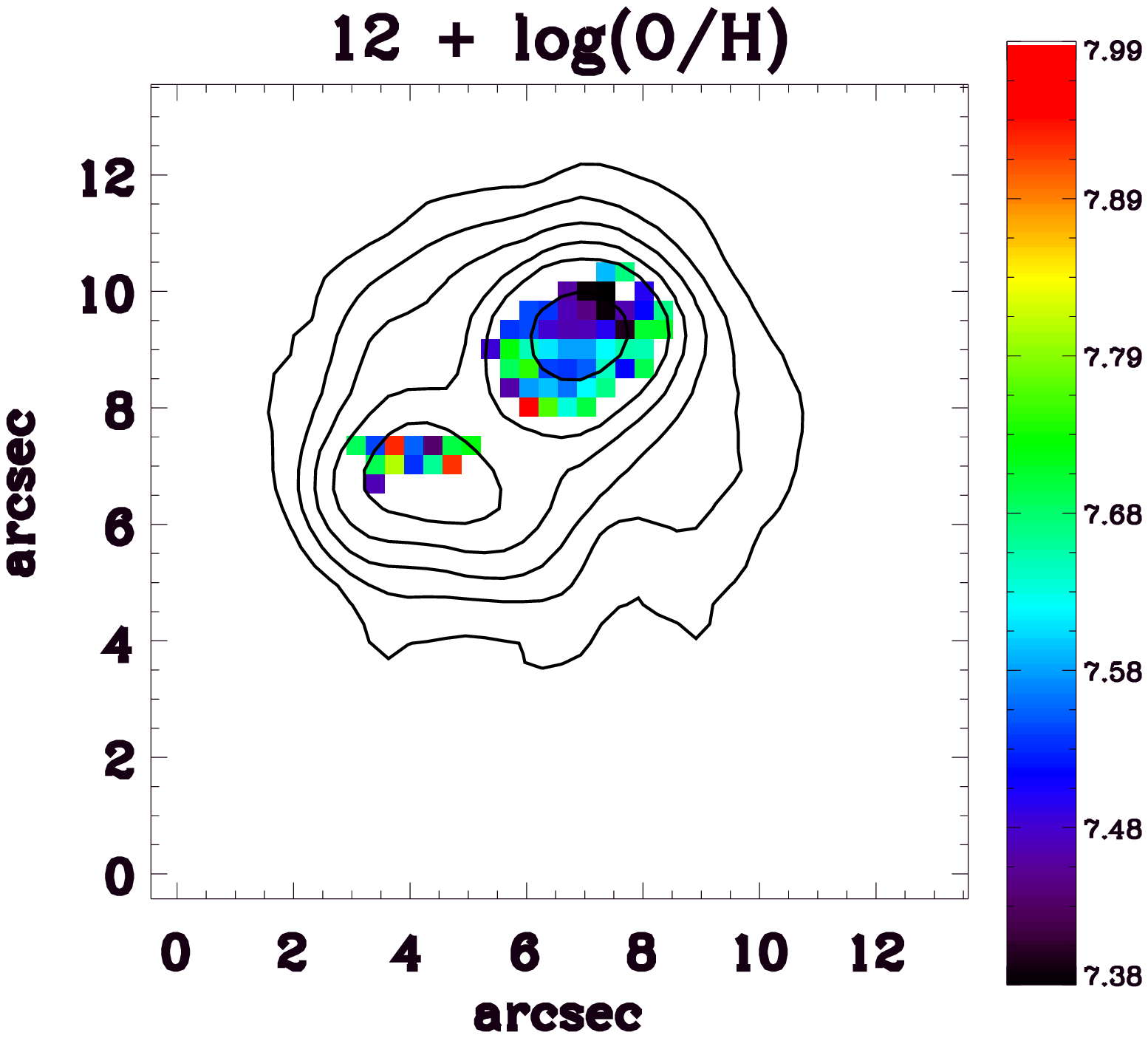}
\includegraphics[width=58mm]{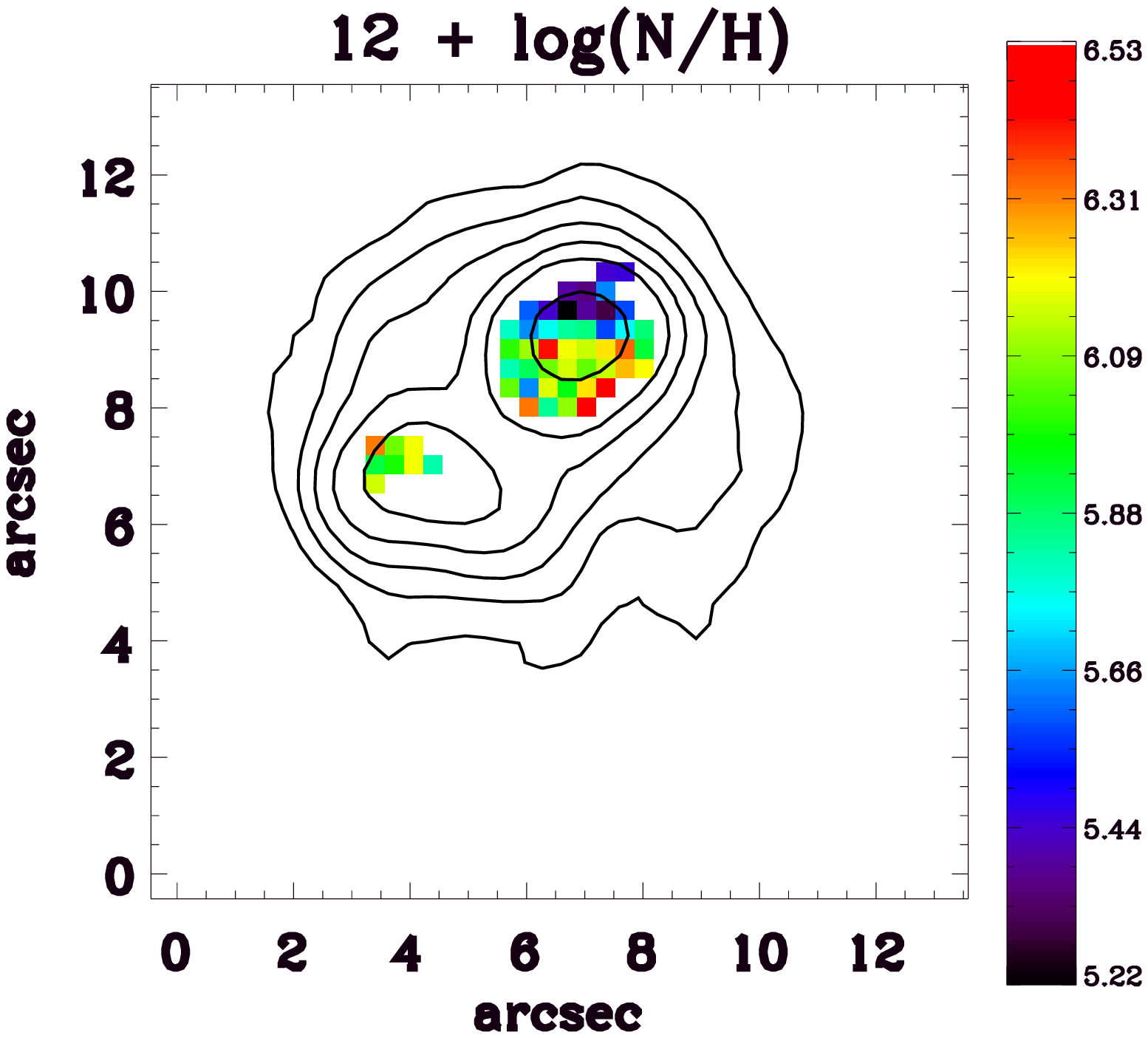}
\includegraphics[width=58mm]{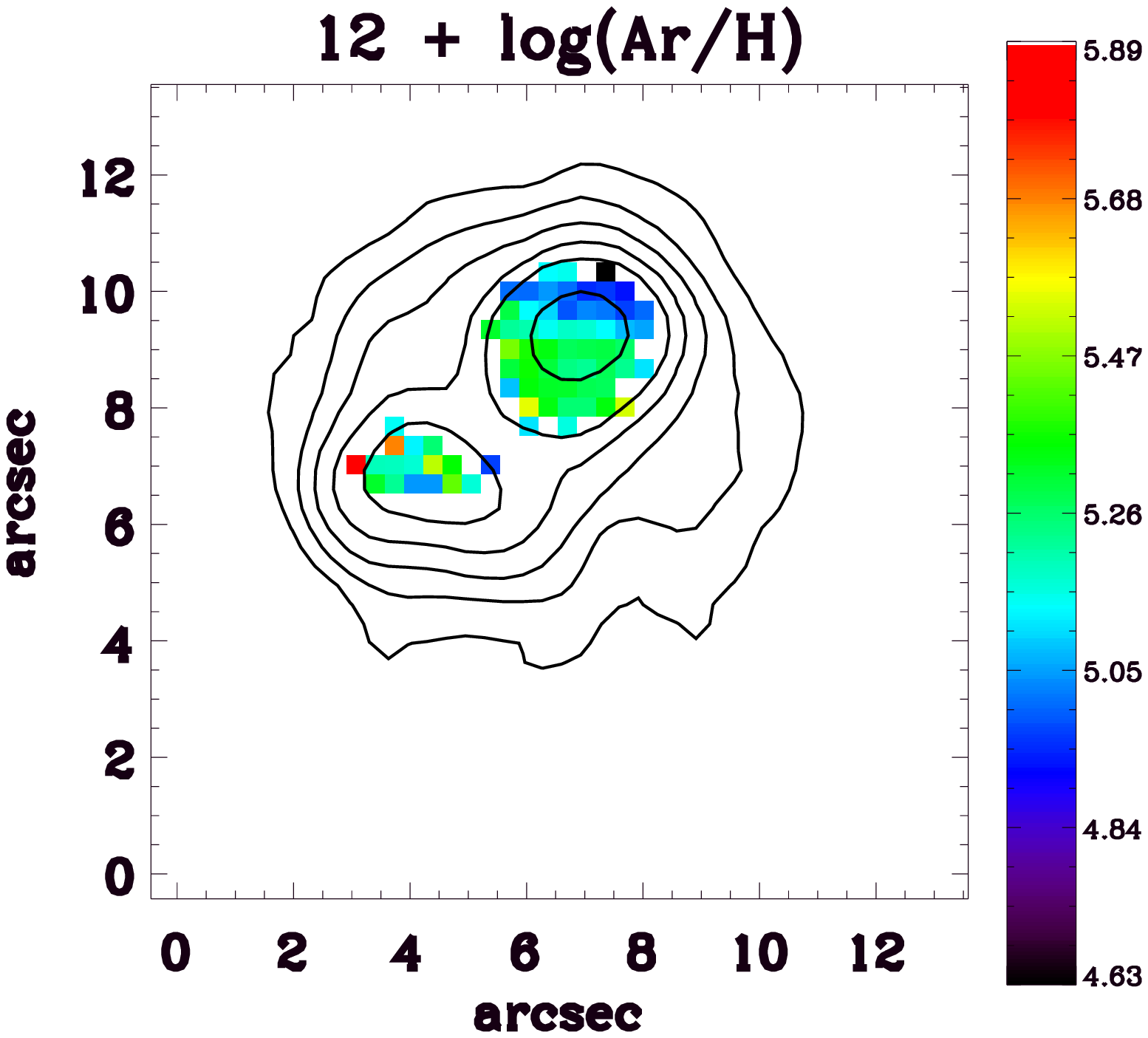}\\
\includegraphics[width=58mm]{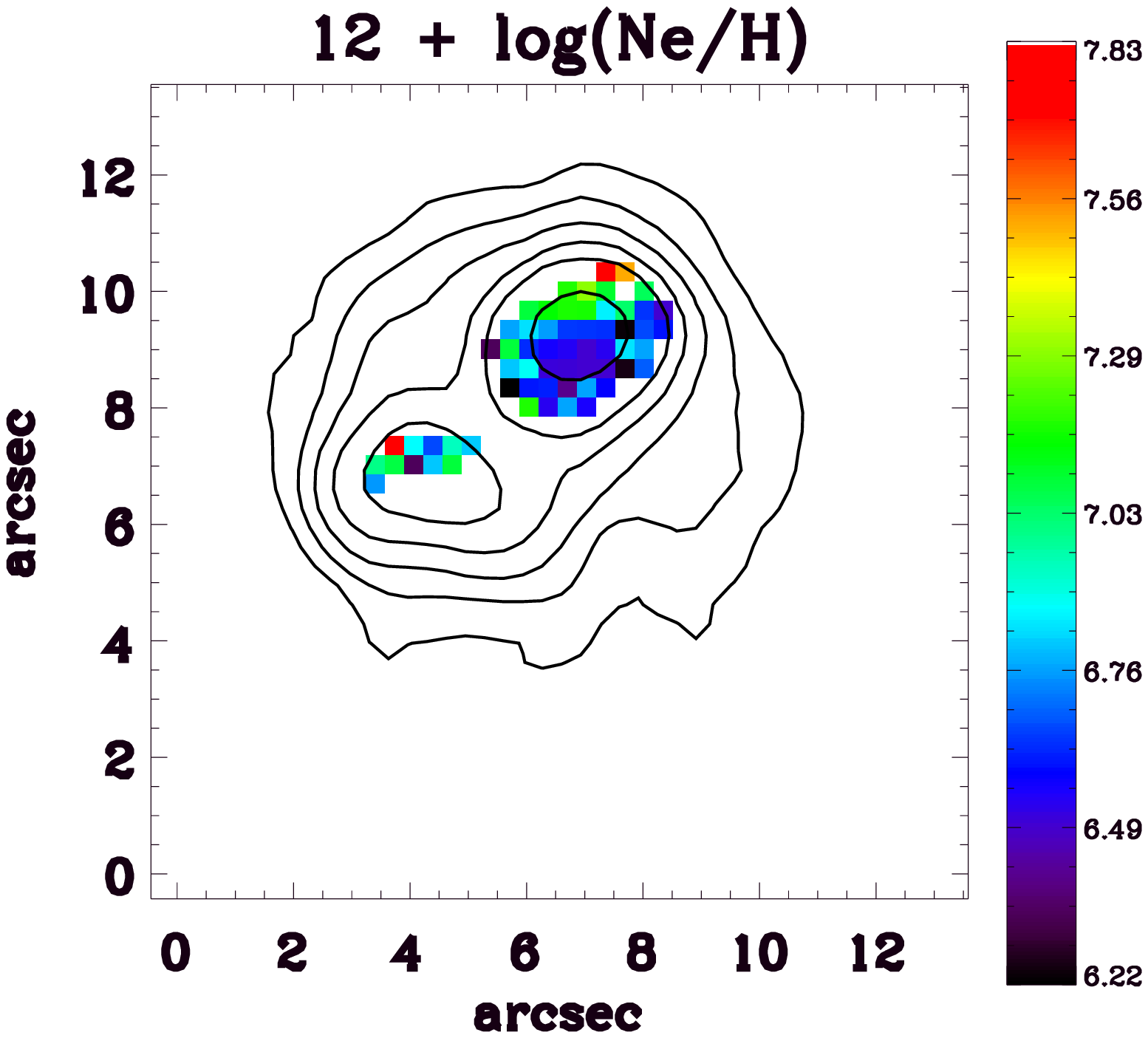}
\includegraphics[width=58mm]{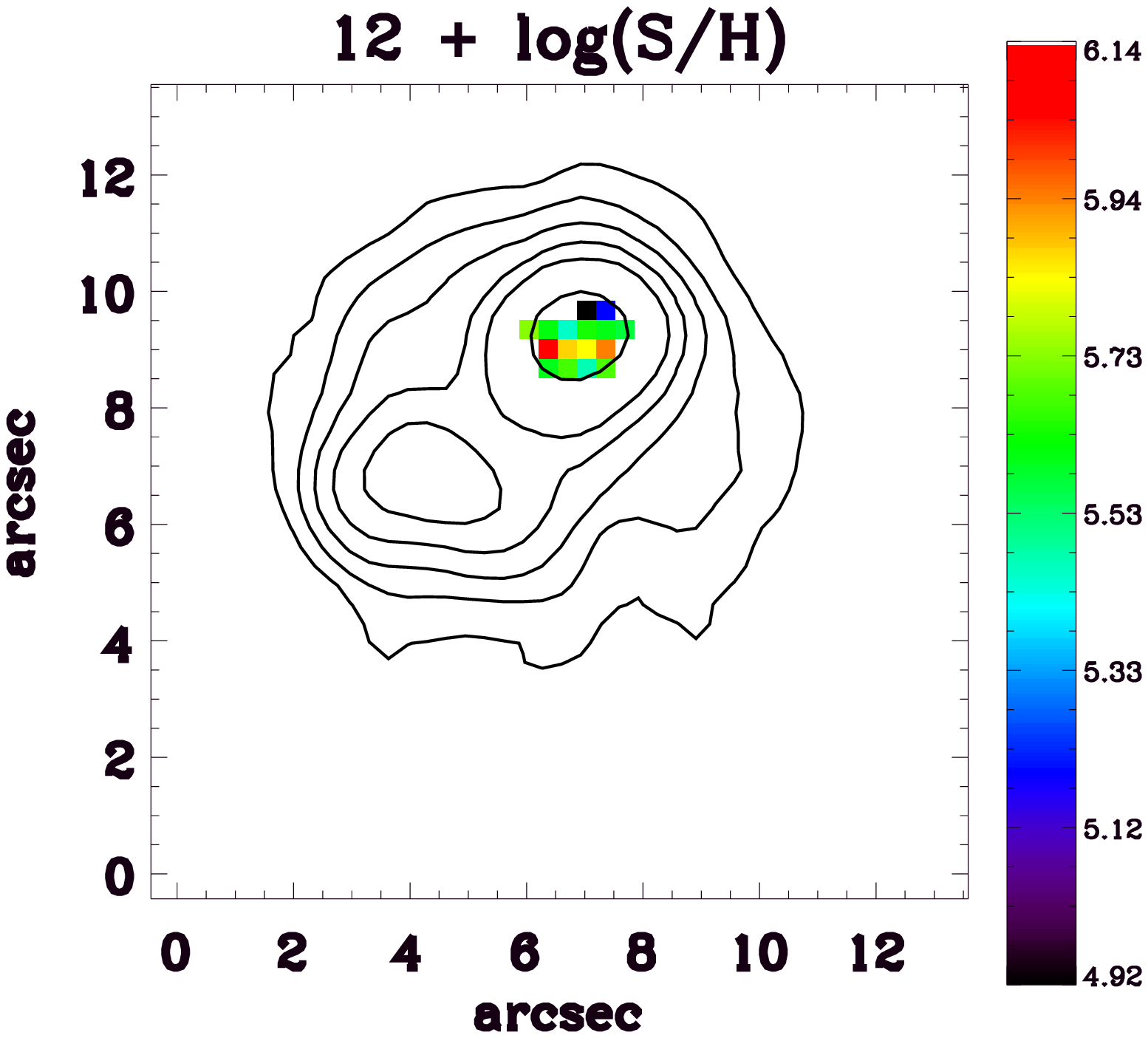}
  \caption{12+log(O/H), 12+log(N/H), 12+log(Ar/H), 12+log(N/H) and 12+log(S/H) abundance maps for Tol 65.
  H$\alpha$ emission line contours are overlaid in the panels. North is up and east is to the left.}
  \label{figure_abundances_Tol65}
\end{figure*}

\begin{itemize}

\item Oxygen: The integrated oxygen abundance of 12+log(O/H)=7.56$\pm$0.06, in the body of the galaxy, is in agreement  
with those provided in the literature, using long-slit, of 7.53$\pm$0.05 \citep{KunthSargent1983}, 
7.59$\pm$0.05 \citep{Pagel1992}, 7.556$^{+0.032}_{-0.035}$ \citep{KunthJourbert1985} and 12+log(O/H)=7.54$\pm$0.01 
\citep{Izotov2001,Izotov2004}. Both GH\,{\sc ii}Rs show the same oxygen abundance, within the uncertainties, with values 
of 12+log(O/H)=7.54$\pm$0.04 and 7.58$\pm$0.08, respectively. 
The 12+log(O/H) values, in Fig. \ref{figure_abundances_Tol65}, range from 7.38 to 7.99 showing 
an almost constant pattern over the GH\,{\sc ii}Rs. 
However, we found that the maximum values of oxygen are not peaked in the centre part of region no. 1, indicating 
a slight rising abundance gradient towards the outskirts of this region.

\item Nitrogen: We found an integrated value for the body of Tol 65 of 12+log(N/H)=5.95$\pm$0.13,
and 5.79$\pm$0.07 and 5.97$\pm$0.18 in regions nos. 1 and 2, respectively.
The nitrogen-to-oxygen ratio in the body is log(N/O)=-1.61$\pm$0.20, and 
-1.75$\pm$0.10 and -1.61$\pm$0.27  in regions nos. 1 and 2, respectively. 
Those values agree, within the uncertainties, with those measured by \cite{Izotov2001} of -1.60$\pm$0.02 
for the integrated galaxy and by \cite{Izotov2004} of -1.595$\pm$0.019 and -1.635$\pm$0.041 
for our regions nos. 1 and 2, respectively.
Therefore, the nitrogen abundances of both star-forming regions, and also the main body, are consistent  
with those of other XMP galaxies in the literature of similar oxygen abundance \citep[see Fig. 13 in][]{Lagos2014}. 
However, there is a slight correlation between the N/O ratio map in region no. 1 and the 
O abundance, in the sense that high N/O values are, preferentially, placed in regions of lower oxygen abundance.
This is compatible by considering the O/H distribution alone without a need to invoke an N enhancement in the nucleus
within the uncertainties.

\item Neon, Argon and Sulfur: The Ne/H, Ar/H and S/H distribution across the GH\,{\sc ii}Rs of Tol 65 are relatively constant, 
within the uncertainties. 
We obtained integrated values of 12+log(Ne/H)=6.66$\pm$0.23, 6.81$\pm$0.12 and 6.74$\pm$0.31, 
12+log(Ar/H)=5.17$\pm$0.05, 5.17$\pm$0.03 and 5.25$\pm$0.06 for the body of the galaxy and regions nos. 1 and 2, respectively. 
For these apertures we found values of log(Ne/O)=-0.90$\pm$0.29, -0.72$\pm$0.15 and -0.83$\pm$0.39;
and log(Ar/O)=-2.39$\pm$0.11, -2.37$\pm$0.06 and -2.33$\pm$0.15, for the same aforementioned apertures.
We derived Sulfur abundances only in region no. 1. Thus, we found values of 
12+log(S/H)=5.67$\pm$0.03 and log(S/O)=-1.86$\pm$0.06 for that region.

\end{itemize}

\begin{figure*}
\includegraphics[width=85mm]{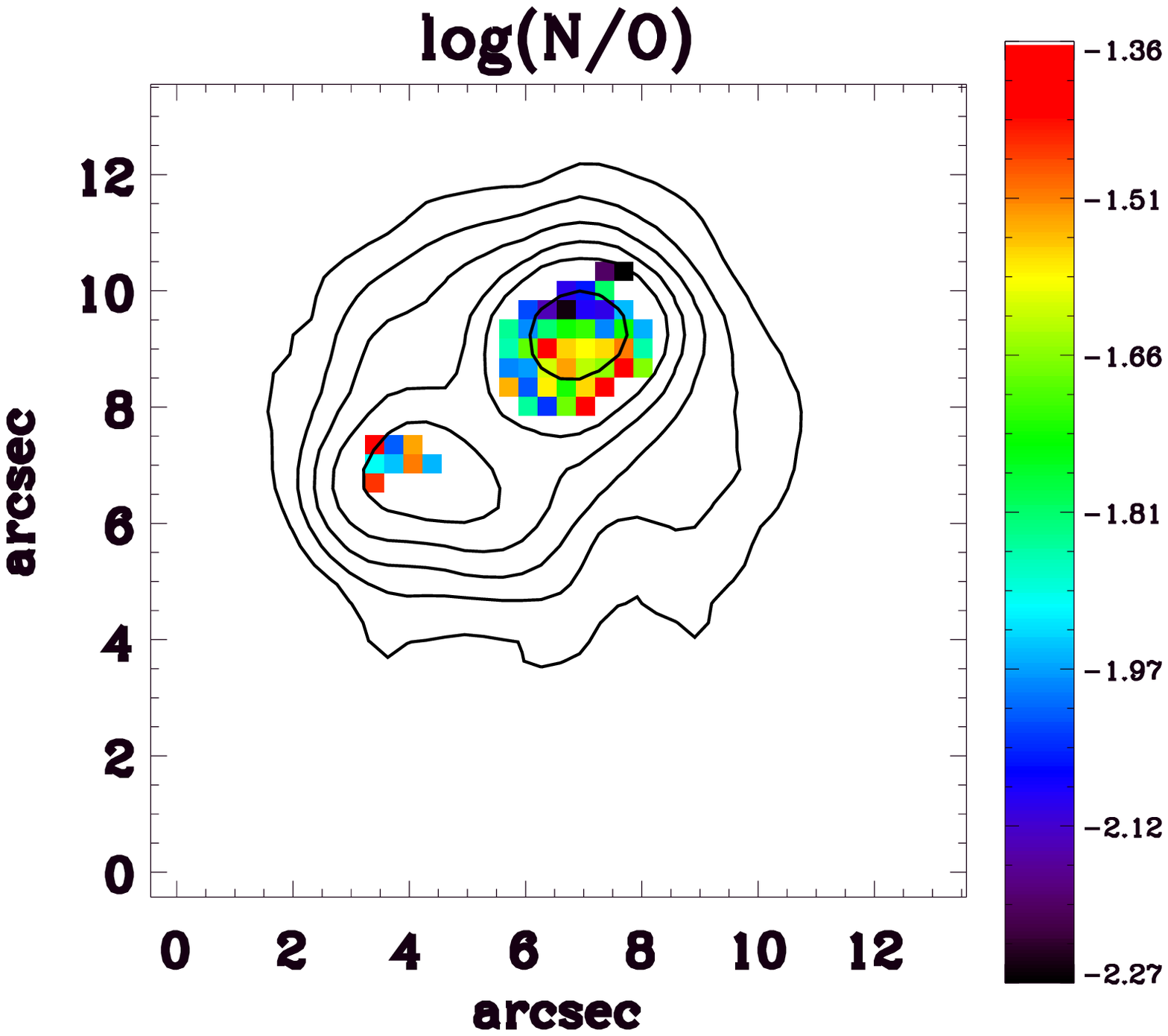}
\includegraphics[width=85mm]{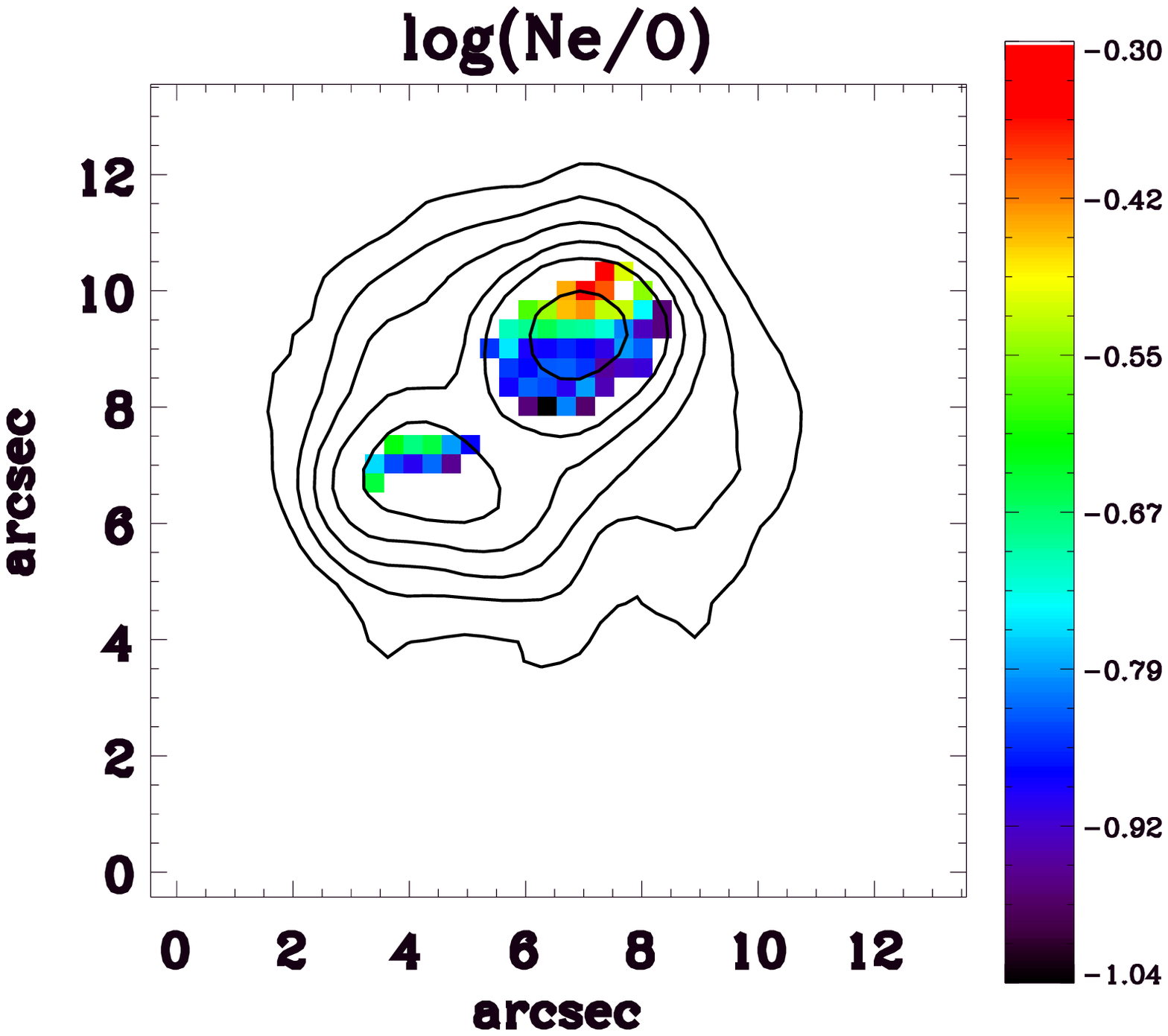}
\includegraphics[width=85mm]{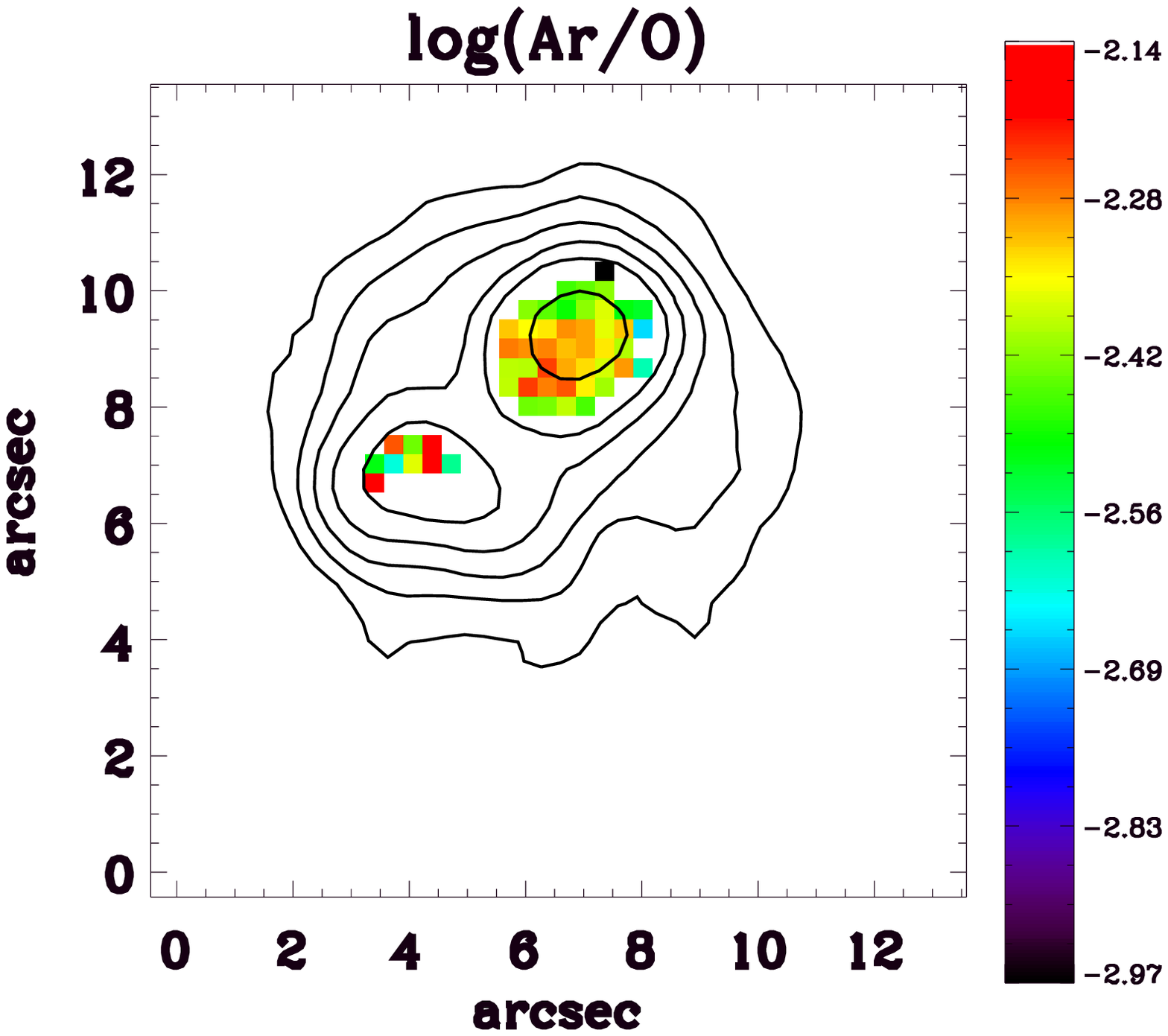}
\includegraphics[width=85mm]{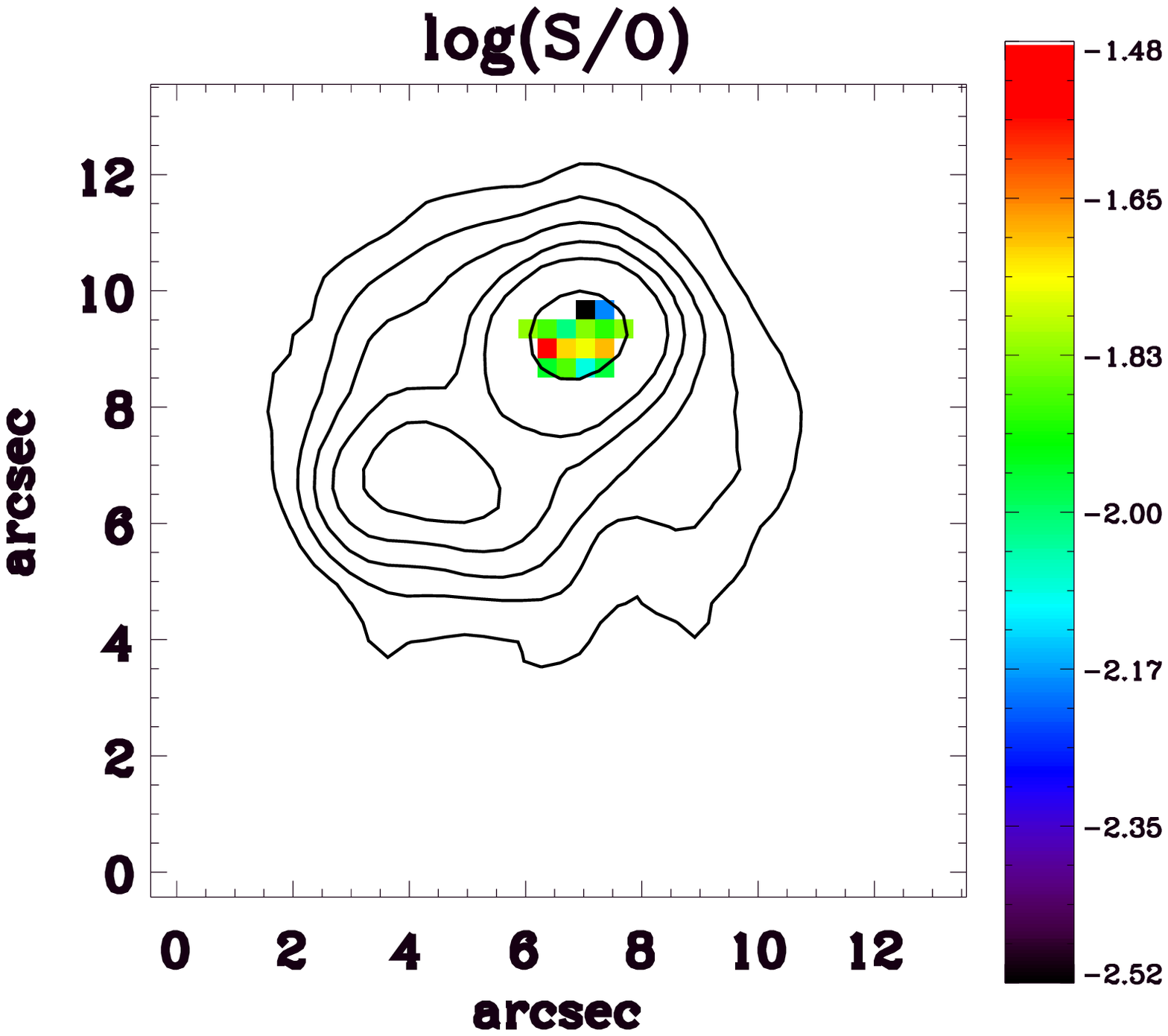}
  \caption{log(N/O), log(Ne/O), log(Ar/O) and log(S/O) abundance maps for Tol 65. 
  H$\alpha$ emission line contours are overlaid in the panels. North is up and east is to the left.}
\label{figure_abundances_ratio_Tol65}
  \end{figure*}

\subsubsection{Summary of the results and comparison with other spaxel scale}\label{binned_spaxels}

\begin{figure}
\centering
\includegraphics[width=85mm]{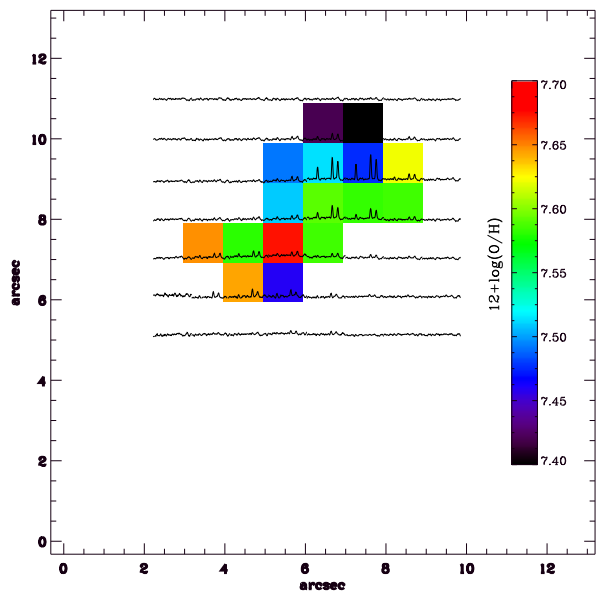}
  \caption{Spatial distribution of the binned ($\sim$1$\arcsec$ spaxels) 12+log(O/H) abundances and 
  [S\,{\sc ii}]$\lambda\lambda$6717,6731 emission line profiles. North is up and east is to the left.}
  \label{figure_sii_profile_properties_Tol65}
  \end{figure}

Figure \ref{figure_sii_profile_properties_Tol65} shows the spatial distribution 
of the binned ($\sim$1.0\arcsec) 12+log(O/H) abundances and [S\,{\sc ii}]$\lambda\lambda$6717,6731 profiles. 
To do this we summed the spaxels, in the data cube, in a region of 3$\times$3 spaxels in order to increase the S/N.
From this figure it is clear that the results obtained previously in this Sect. are 
consistent when we binned the spaxels from the 0.33$\arcsec$ to $\sim$1.0$\arcsec$, which corresponds
approximately with the mean value of the seeing during the observations. 
We note in Fig. \ref{figure_sii_profile_properties_Tol65} that the intensity of the central peak 
of the [S\,{\sc ii}]$\lambda$6731 emission line is slightly higher than the intensity of 
[S\,{\sc ii}]$\lambda$6717 in a few spaxels in between the GH\,{\sc ii}Rs. 
This indicate that this area shows high electron density values $\gg$100 cm$^{-3}$. 
On the other hand, we did not observe any chemical inhomogeneity, within the uncertainties, along 
the main body of the galaxy. However, the highest values of 12+log(O/H) and lowest values of t$_e$(O\,{\sc iii}) 
are placed on the southern part of region no. 1 and following the same pattern 
found at 0.33$\arcsec$ spatial scale, while the 12+log(O/H) for the different apertures considered here are similar. 
Therefore, the spatial distribution of O, Ne, S, Ar and also N/O, Ne/O, Ar/O and S/O abundances 
in this XMP H\,{\sc ii}/BCD appear to be flat at large scales. 
This result agrees with previous studies in the literature 
\cite[e.g.][and references therein]{Lagos2009,Lagos2012,LagosPapaderos2013} in the sense that
no spatial variations of oxygen abundance has been found in H\,{\sc ii}/BCD galaxies.
Finally, it is interesting to note that if we assume an electron density of n$_{e}\sim$100 cm$^{-3}$ 
for the individual spaxels we obtain a mean abundance value of 7.50 in units of 12+log(O/H). 
Meanwhile if we assume a density of 1000 cm$^{-3}$ we obtain that the mean value of 12+log(O/H)=7.63. 
Therefore, no significant changes in metallicity are found, within the uncertainties, when we assume a constant 
density of 100 cm$^{-3}$ through the ISM of the galaxy.

\subsection{Gas Kinematics}\label{sect_velocity_field}

\begin{figure*}
\includegraphics[width=85mm]{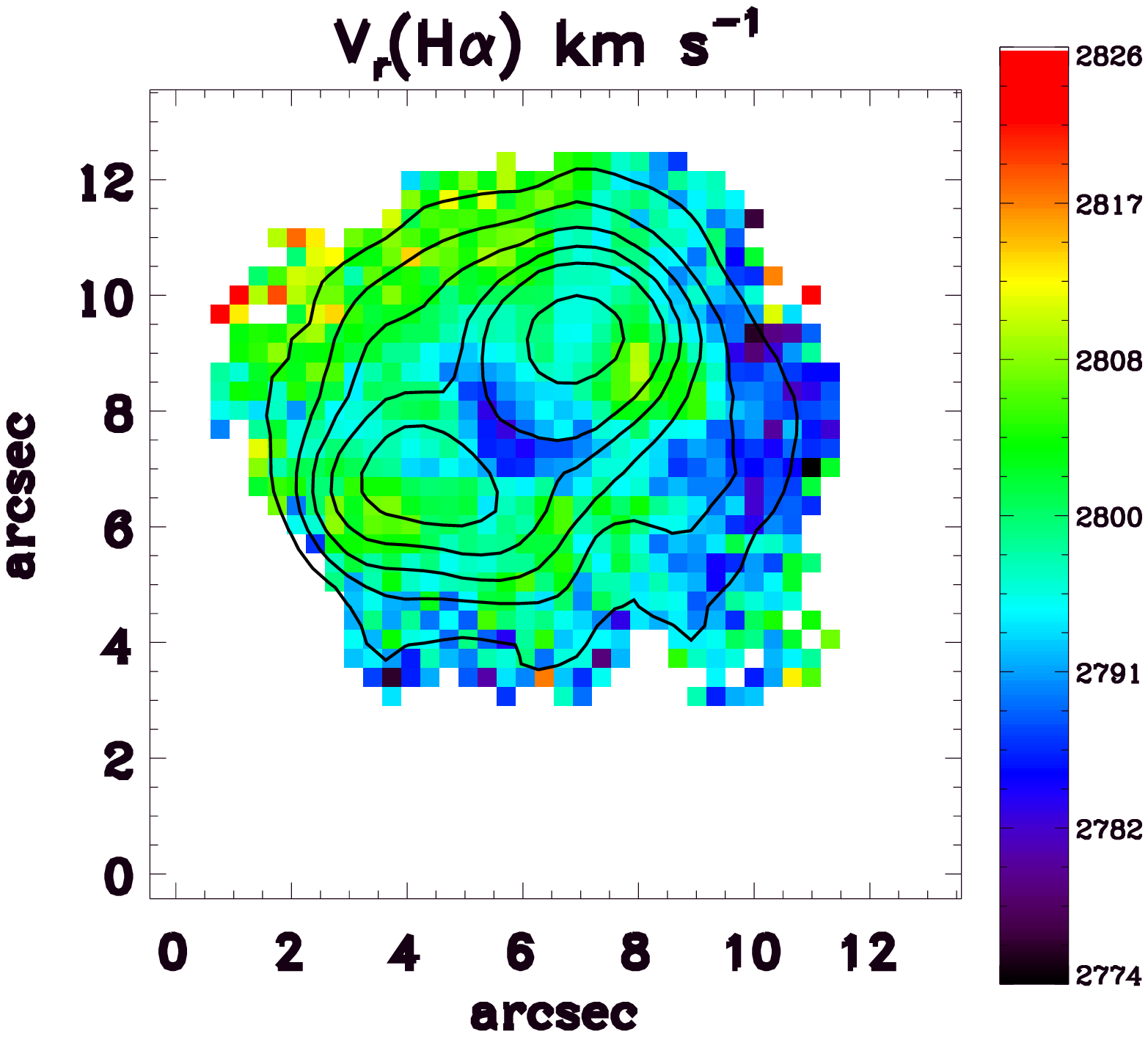}
\includegraphics[width=85mm]{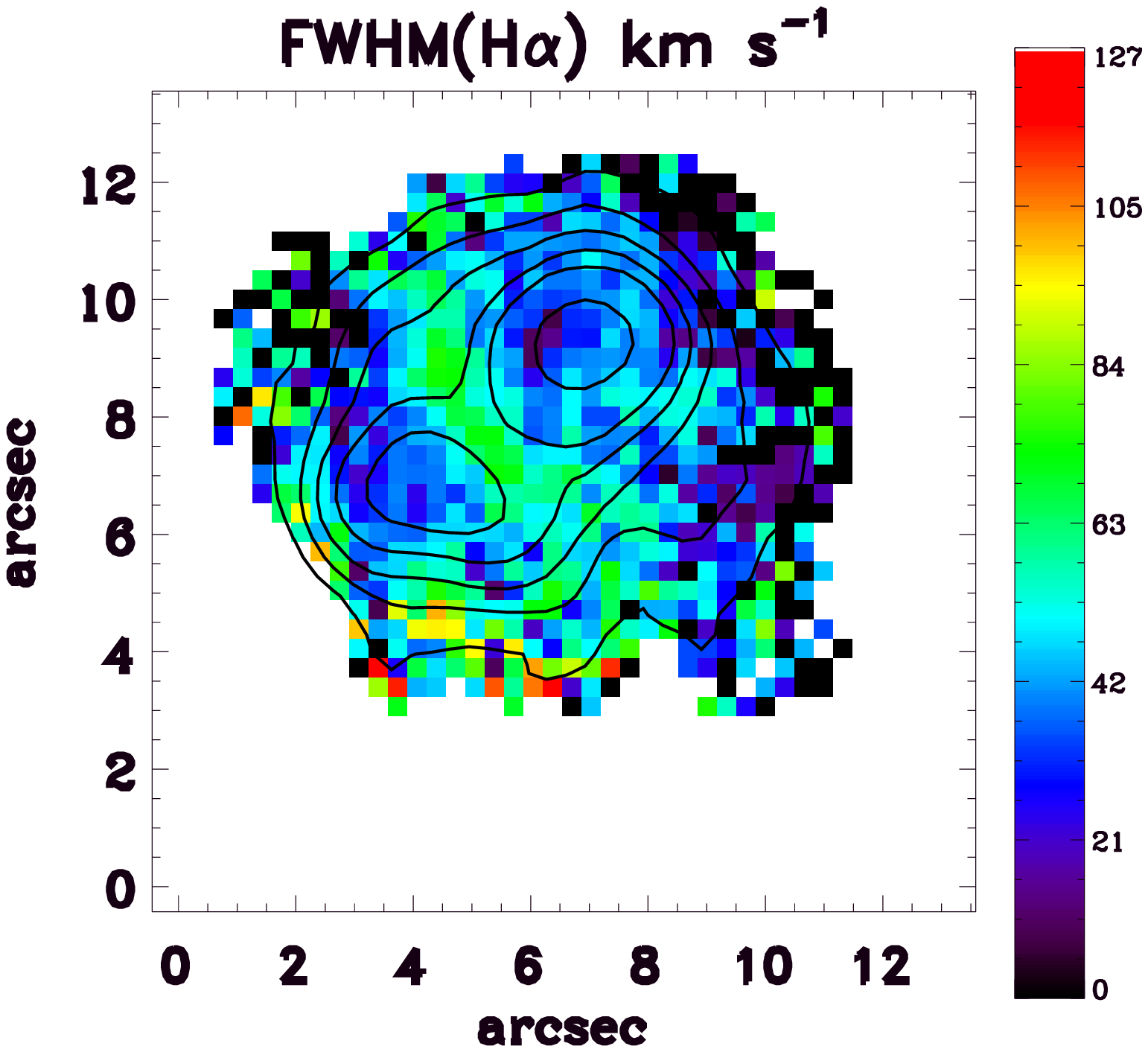}
  \caption{Radial velocity and FWHM maps of the ionized gas obtained from the H$\alpha$ emission line.
  FWHM corrected by instrumental and thermal broadening. H$\alpha$ emission line contours are overlaid in the panels.
  North is up and east is to the left.}
\label{figure_velocity_field_Tol65}
  \end{figure*}

The kinematics distribution of the emitting gas was obtained by fitting Gaussian curves to the line profiles of H$\alpha$.
The complexity of the internal motions in Tol 65 can be seen in Fig. \ref{figure_velocity_field_Tol65}.
The radial velocity V$_r$(H$\alpha$) map of the galaxy (left-hand panel) shows no spatial correlation 
with the H$\alpha$ emission and we found no evidence of velocity gradients or rotational patterns. 
Meanwhile, the FWHM(H$\alpha$) map (right-hand panel) shows the lowest values 
on the two star-formation regions. We note that these velocity fields are akin the ones observed in \,{\sc ii} Zw 40
by \cite{Bordalo2009}, in the sense that some regions does not seem to play a significant role stirring the gas up, 
then likely preserving the kinematic signature of the proto-cloud, while extended regions of the low intensity and diffuse gas 
are highly disturbed, probably due to unresolved expanding shells and the effects of massive star evolution 
(i.e., supernovae, champagne flows, relative motions between discrete clusters providing large scale shear, and turbulence and rotation).

Despite the different sizes, the two main regions have the same kinematic properties, with very similar 
patterns. Both regions have smooth line profiles with $\sigma\sim$19.6$\pm$0.6 km s$^{-1}$ and 22.5$\pm$0.4 km s$^{-1}$  
and a heliocentric V$_r\sim$2805.51  km s$^{-1}$ and  2808.03 km s$^{-1}$, respectively. 
The velocity dispersion ($\sigma$=FWHM/2.35) was obtained considering 
the instrumental dispersion, see Sect. \ref{sect_obser_reduc}, and $\sigma_{th}$=$\sqrt{kT_e/m_H} \sim$12 km s$^{-1}$
for the thermal broadening at t$_e$(O \,{\sc iii})=17087 K, thus $\sigma^2$=$\sigma^2_{obs}$-$\sigma^2_{inst}$-$\sigma^2_{th}$.
Our corrected velocity dispersion $\sigma$(H$\alpha$)=20.8$\pm$0.6 km s$^{-1}$ for the body of the galaxy
agrees with the value obtained by \cite{BordaloTelles2011} of $\sigma$(H$\alpha$)=18.4$\pm$0.4 km s$^{-1}$.
This is identical, within the errors, to the velocity of the H\,{\sc i} cloud with a line width of $w_{50}$=40$\pm$6 km s$^{-1}$
($\sigma \sim$17 km s$^{-1}$) obtained by \cite{PustilnikMartin2007}. This implies that the ionized gas still retains 
the kinematic memory of its parental cloud likely to be associated with the gravitational
potential well \citep[see][]{Bordalo2009}.

It is interesting to note that the zone in between regions nos. 1 and 2, see Fig. \ref{figure_velocity_field_Tol65}, 
shows low V$_r$(H$\alpha$) and high FWHM(H$\alpha$) values. This suggest that
the kinematic structure of the galaxy is likely dominated by the current burst of star-formation
\citep[e.g.][]{Moiseev2015}. It may indicate a weak outflow seen towards the observer.
Therefore, turbulent motions resulting from the starburst appear to dominate over rotation in this galaxy.
\cite{Moiseev2015} obtain a similar conclusion based in the analysis of a large sample of H\,{\sc ii}/BCD 
galaxies obtained using Fabry-Perot interferometry.  
In addition, we found a velocity dispersion slightly lower in the tail of the galaxy with a value of 
$\sigma$(H$\alpha$)=17.5$\pm$1.0 km s$^{-1}$. The difficulties in calculating these values and the small difference
within them lead us to consider $\sigma_{body}\simeq\sigma_{tail}$.
Finally, we did not observe the presence of asymmetric line profiles or/and
a broad component in the base of the emission lines.
We will discuss in detail in Sect. \ref{sect_discussion} 
the results obtained in this section and its implications in the evolutionary stage of the galaxy.

\subsection{Mass of the ionized hydrogen in the tail of the galaxy}\label{sect_tail_gasmass}

\begin{figure}
\includegraphics[width=85mm]{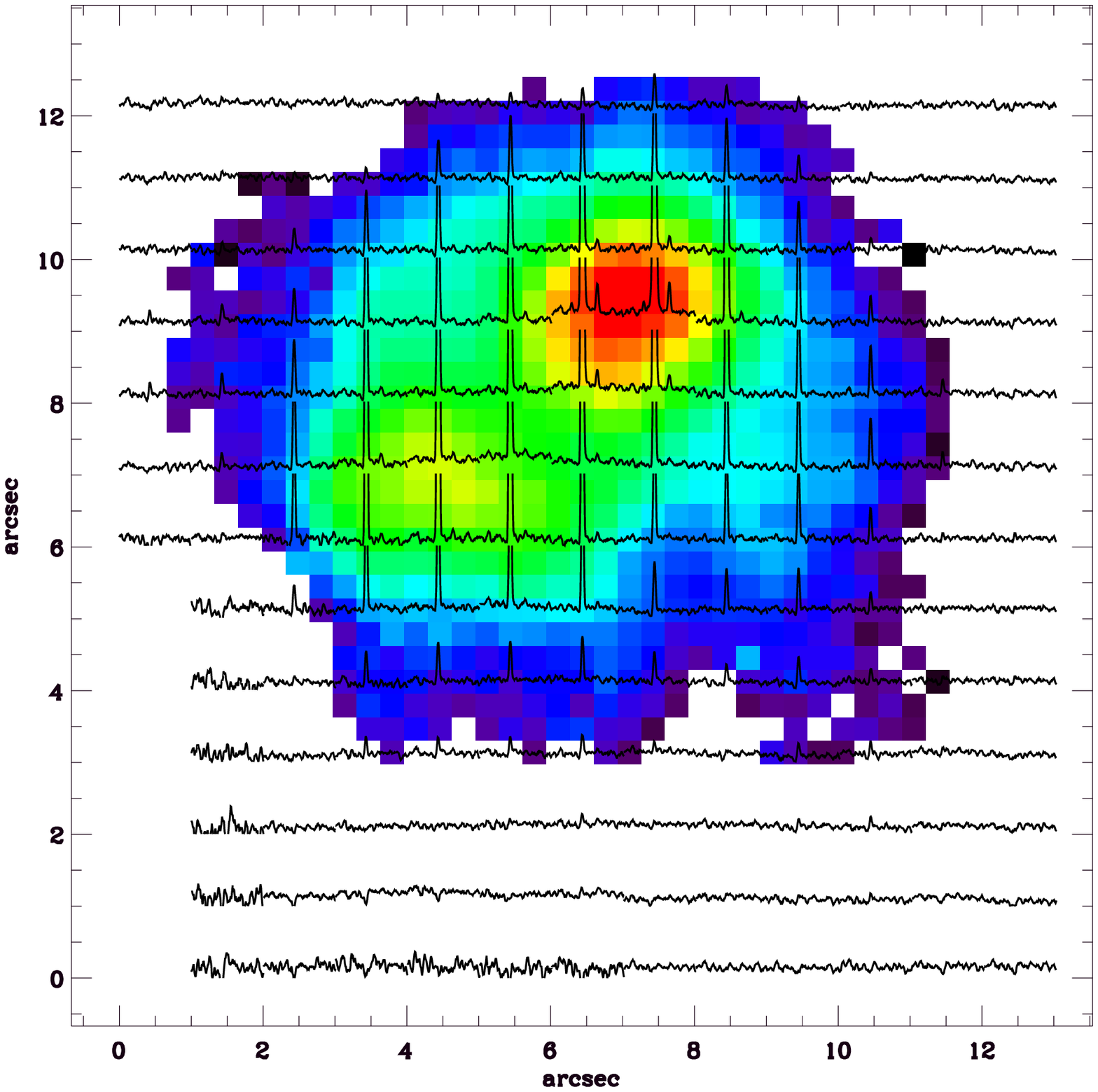}
\includegraphics[width=85mm]{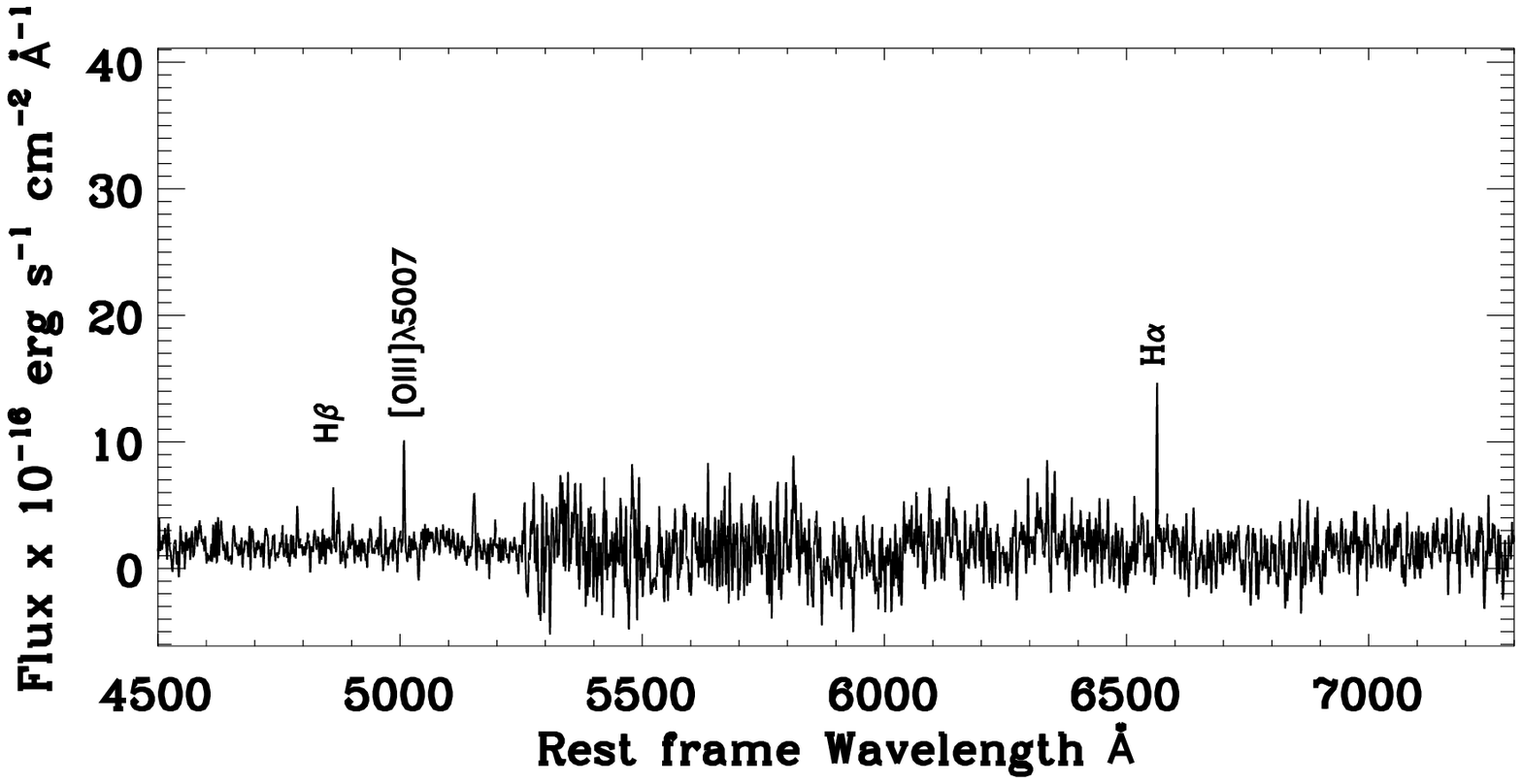}
  \caption{Upper panel: Spatial distribution of the binned ($\sim$1$\arcsec$ spaxels) H$\alpha$ profiles 
  in the galaxy over the 0.33$\arcsec$ spaxels. See Sect. \ref{binned_spaxels}. North is up and east is to the left.
  Lower panel: Integrated spectrum of the tail. From this fig. we observe the presence of an extended nebular 
  emission in the tail of the galaxy.}
  \label{figure_emission_tail_Tol65}
\end{figure}

In the upper panel of Fig. \ref{figure_emission_tail_Tol65} 
we show the spatial distribution of the binned ($\sim$1$\arcsec$) H$\alpha$ emission line profiles, 
while in the lower panel of the same Fig. we show the integrated spectrum of the tail. 
In Table \ref{table_flux_tail} we show the fluxes and the extinction obtained in that region. 
It is clear from Fig. \ref{figure_emission_tail_Tol65} and also Fig. \ref{figure_images_ha} (right-hand panel)  
that there is an extended nebular emission in the tail of the galaxy. 
The presence of this extended nebular emission can have a large impact on the inferred properties of the underlying stellar
component \cite[e.g.][]{PapaderosOstlin2012}. 
According to \cite{Duc2004} the effect of tidal perturbations 
can efficiently carry away from the disk a large fraction of the gas. 
At this point, we can ask if the observed ionized gas in the tail of the galaxy 
is it the result of a tidal perturbation, infall of gas or merely the consequence of the star-formation activity
(see Sect. \ref{sect_disc_origin}).

\begin{table}
\centering
 \begin{minipage}{60mm}
  \caption{Emission lines and physical properties of the gas in the tail of the galaxy. 
The fluxes are relative to F(H$\beta$)=100.}
  \label{table_flux_tail}
  \begin{tabular}{@{}lccccccccr@{}}
  \hline
      &    \\
      &  F($\lambda$)/F(H$\beta$)  & I($\lambda$)/I(H$\beta$)         \\ 
 \hline
$\left[OIII\right] \lambda$5007   & 227.38$\pm$2.81 &  222.88$\pm$3.90\\
H$\alpha \lambda$6563             & 330.30$\pm$7.19 &  282.14$\pm$8.68\\
                                  &                                   \\
F(H$\beta$)\footnote{In units of erg cm$^{-2}$ s$^{-1}$} & (1.16$\pm$0.01)$\times$10$^{-15}$\\
c(H$\beta$)                       & 0.23                              \\
EW(H$\beta$)\footnote{$\rm \AA$}  & 7                                 \\
\hline

\end{tabular}
\end{minipage}
\end{table}

In this context, an estimate of the mass of emitting ionized hydrogen in the tail of the galaxy, 
is an important indicator of the amount of material present in this structure. This can be made using 
the H$\alpha$ emission-line luminosity L(H$\alpha$)=7.14$\times$10$^{38}$ erg s$^{-1}$ 
and the expression for the mass of the ionized gas from \cite{OsterbrockFerland2006}; 

\begin{equation}
M(HII)=\frac{L(H\alpha)m_{p}}{n_{e}\alpha_{H\alpha}^{eff}h\nu_{H\alpha}},
\end{equation}

where m$_p$ is the mass of the proton and n$_e$ is the electron density.
We use the effective recombination coefficient $\alpha^{eff}$
for H$\alpha$ and case B emission at T=2$\times$10$^4$ K \citep{OsterbrockFerland2006}.
Thus, assuming an average electron density of $\sim$10 cm$^{-3}$, we find a total mass 
of ionized hydrogen in the tail of M(H\,{\sc ii})$\sim$1.70$\times$10$^5$ M$_\odot$.
This mass corresponds with $\sim$24 per cent of the total ionized mass\footnote{
We used the total H$\alpha$ flux of the galaxy summing all the spaxels over the FoV.
Thus, F(H$\alpha$)/F(H$\beta$)=3.4899 with F(H$\beta$)=47.48$\times$10$^{-15}$
erg cm$^{-2}$ s$^{-1}$.} of the galaxy (or $\sim$3 per cent if we consider $\sim$100 cm$^{-3}$) 
with M(H\,{\sc ii})$\sim$7.00$\times$10$^5$ M$_\odot$. 
Although this mass estimate is approximate, it could clearly indicate that there is a large 
amount of H\,{\sc i} in this extended structure.

\section{Discussion}\label{sect_discussion}

\subsection{Star-formation rate and metallicity}

The star-formation rate (SFR) inferred using the \cite{Kennicutt1998} formula 
after correction for a Kroupa IMF \citep{Calzetti2007}, is 0.152 M$_{\odot}$yr$^{-1}$ 
for the main body of the galaxy and 0.090 M$_{\odot}$yr$^{-1}$ 
and 0.016 M$_{\odot}$yr$^{-1}$ for regions nos. 1 and 2, respectively.
This global SFRs are low in absolute terms. However, its starbursting nature appears 
when one computes the specific SFR, i.e. the SFR per unit of mass. Somewhat equivalently,
their SFR per unit of area ($\Sigma_{SFR}$). We can translate the integrated SFR of the main body into an integrated 
$\Sigma_{SFR}$ assuming an aperture of $\sim$1.95 Kpc$^2$ that corresponds
to the total area of this region. 
Thus, we obtain that $\Sigma_{SFR,body}\simeq$0.078 M$_{\odot}$ yr$^{-1}$kpc$^{-2}$.
If we compare this result with the one obtained in our previously analyzed 
XMP BCD galaxy, using IFU spectroscopy, \citep[HS 2236+1344,][]{Lagos2014} we found that in Tol 65 the $\Sigma_{SFR}$
is 2 times lower, while the $\Sigma_{SFR,body}$ is comparable to the star-formation 
per unit of area of local tadpole galaxies \citep{Elmegreen2012}.
On the other hand, the $\Sigma_{SFR}$ in Tol 65, and also in HS 2236+1344, is above the SFRs found in 
``normal" or more metal rich H\,{\sc ii}/BCD galaxies. From Fig. 15 in \cite{LopezSanchez2010}
it is evident that the SFR per unit of area in our XMP galaxies studied so far are higher than the ones found in 
most of their sample of H\,{\sc ii} galaxies. This is explained by the high concentration of H\,{\sc i} gas found in XMPs
\citep[e.g.][]{EktaChengalur2010b,Filho2013}.

\begin{figure}
\includegraphics[width=85mm]{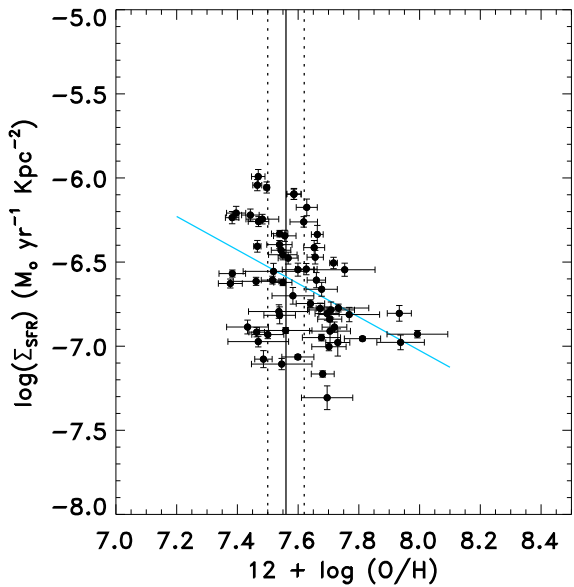}
  \caption{Relation between the star-formation rate surface density $\Sigma_{SFR}$ and oxygen abundance 12+log(O/H) 
  at spaxel scales. The cyan line represents the linear fit to this relation. 
  The 12+log(O/H)=7.56 obtained in the main body of the galaxy is represented by the continuous line, while the errors 
  at 1$\sigma$ level are in dotted lines.}
  \label{figure_SFR_OH}
\end{figure}

Using the results obtained in Sect. \ref{sect_abundances} we show, in Fig. \ref{figure_SFR_OH}, 
the spatially resolved relation between the $\Sigma_{SFR}$ and the oxygen abundance 12+log(O/H) in Tol 65. 
Even this relation at spaxel scales is dubious, 
our findings show a marginal anticorrelation between these quantities in the sense that high spatial star-formation is found 
in regions of lower metallicities (see the linear fit to this relation in Fig \ref{figure_SFR_OH}). 
The calculation of Pearson's correlation for these data gives a value of -0.41,
that indicate a moderate monotonically decreasing relationship between these quantities.
If we interpret this result as the consequence of an ongoing infall 
of metal-poor gas from the outskirts the scatter ($\sigma_{(O/H)}$=0.13 dex) in Fig. \ref{figure_SFR_OH} 
clearly indicates that the metals in the ISM are almost fully diluted. 
Therefore, from an statistical point of view the ISM of the galaxy can be considered 
chemically homogeneous. We will come back on this discussion later in Sect. \ref{sect_disc_properties}. 

\subsection{Relation between the starburst activity and the Ly$\alpha$ absorption}\label{sect_starburst_Lalpha}

It is however worth noting that the age ($\sim$3-5 Myr; see Sect. \ref{sect_I_c_EW}) of the current burst of Tol 65 agrees  
with the idea proposed by \cite{ThuanIzotov1997}, in the sense that this galaxy is a young starburst embedded 
in a static H\,{\sc i} cloud, which produces a damped Ly$\alpha$ absorption.
\cite{ThuanIzotov1997} found even with the poor quality of their spectrum, that the O\,{\sc i} $\lambda$1302
absorption line in Tol 65 is blue shifted by $\sim$200 km s$^{-1}$ with respect
to the emission lines and the Si\,{\sc ii}$\lambda$1304 absorption line. 
They argue that if this is true, the O\,{\sc i} absorption would be produced in gas shells moving outward 
from the central star clusters \citep[][]{Lequeux1995,Kunth1998}.
The spatial distribution of electron density n$_e$(S\,{\sc ii}), in Fig. \ref{figure_temp_den_galaxies}, shows a clear
area of constant values in region no. 1 with n$_e$(S\,{\sc ii})$\sim$200 cm$^{-3}$ surrounded by relatively
higher density values $>$200 cm$^{-3}$. In particular, the area in between the two regions 
shows spaxels which reach values of n$_e$(S\,{\sc ii})$\sim$500 cm$^{-3}$ and 
the [S\,{\sc ii}]$\lambda\lambda$6717,6731/H$\alpha$ ratio (see Fig. \ref{figure_ratios_galaxies}) is slightly enhanced in this  
inter-cluster region which would be consistent with an enhanced contribution to the ionization/excitation by shocks.
Therefore, the gas ejected by massive stars within the star clusters grows close to the centre
creating a high density region at the same time that the gas expands with velocities close or higher than the supersonic values 
(with $\sigma >$10 km s$^{-1}$) into the surrounding region, then compressing the gas into high-density condensations 
and creating a low density region between the centre and the expanding high density shells. 
We found high FWHM values in the same area of high electron density in between the GH\,{\sc ii}Rs 
(see Fig. \ref{figure_velocity_field_Tol65}) indicating that the turbulent motions in the ionized gas could be    
produced by winds from the main star cluster complexes. Likewise, the expanding gas would cause the expansion 
in the surrounding material (H\,{\sc i}), so moving it out of the galactic plane.
This agrees with the arguments proposed by \cite{ThuanIzotov1997}
and the evolutionary models by \cite{TenorioTagle1999} in order to produce a damped Ly$\alpha$ absorption.
In addition, if the wind material contains enough dust, this outflow of gas would give rise to blueshifted velocities 
(compared to the clusters) observed in between the GH\,{\sc ii}Rs because redshifted emission 
would be preferentially extinguished by the dust. 
In fact, the lack of Ly$\alpha$ emission \citep[see Fig. 2 in][]{Atek2008} and the detection of relatively high c(H$\beta$)
associated with the most intense star-forming region in this galaxy, region no. 1, suggest
that the Ly$\alpha$ photons have likely been destroyed by dust. This is in good agreement with previous findings 
by \cite{Atek2008} in the sense that dust extinction is an important Ly$\alpha$ escape regulator. 

\subsection{The cometary morphology of Tol 65}\label{sect_disc_origin}

\subsubsection{Feedback from the star-formation activity}

A number of recent studies \citep[e.g.][]{Papaderos2008,ElmegreenElmegreen2010,Elmegreen2012} have suggested 
that local galaxies with cometary or tadpole morphology could have a variety of origins. 
\cite{Elmegreen2012} suggest that most local tadpoles are bulge-free galaxy disks with lopsided star-formation, 
likely from environmental effects (e.g, ram pressure, disk impacts or random collapse of local disk 
gas with an unstable Jeans length).
Alternatively, propagating star-formation along the tail has been proposed by \cite{Papaderos2008} 
as one of the mechanisms that lead to the formation of the elongated underlying component or stellar tail 
of cometary H\,{\sc ii}/BCDs. Given that the starburst produces enough energy from stellar winds and supernovae (SNe),  
could the stellar feedback from the starburst have created the stellar tail in this galaxy? 

To create a $\sim$1.5 kpc long extension in $\sim$5 Myr (age of the current burst) would require an outflow velocity 
$\sim$293 km s$^{-1}$, while the difference between the velocities found in the field is 
$\sim$ 50 km s$^{-1}$. Therefore, the current star-formation activity appears to be too young to have pulled out
the surrounding gas and stars, driven a galactic outflow, to form the tail. On the other hand,
if the global starburst take longer $\sim$10$^8$ yr \cite[e.g.][]{McQuinn2009} as compared to the current burst, 
certainly it may promote the formation of gas outflows. 
After that some part of the gas could decelerate and collapse to form the stellar tail.
If so, the close similarity between $\sigma_{body}$ and $\sigma_{tail}$ agrees with this scenario, given that  
the diffuse gas will acquire the kinematical information of the system which is transferred into the ISM
by the star-formation activity \citep[e.g.][]{Moiseev2015}.  
\cite{Papaderos1999} found that the relatively red colors of the underlying stellar component, in Tol 65, 
suggest a stellar population with a mean age of $\la$1 Gyr. This stellar age is compatible with this idea, 
but the intrinsic youth of XMPs is questionable \citep{Papaderos2008} and the evidence of such strong winds and  
the presence of extended ($\ga$1 kpc) super-shells in low luminosity dwarf galaxies is also sparse.
Finally, we did not observe evidences of propagating star-formation along the galaxy 
and the presence of star clusters in the tail. Thus, there seems to be no correlation between the star-forming activity 
and the formation of the tail in this galaxy, but from our present data we cannot rule out completely this possibility. 

\begin{figure}
\includegraphics[width=80mm]{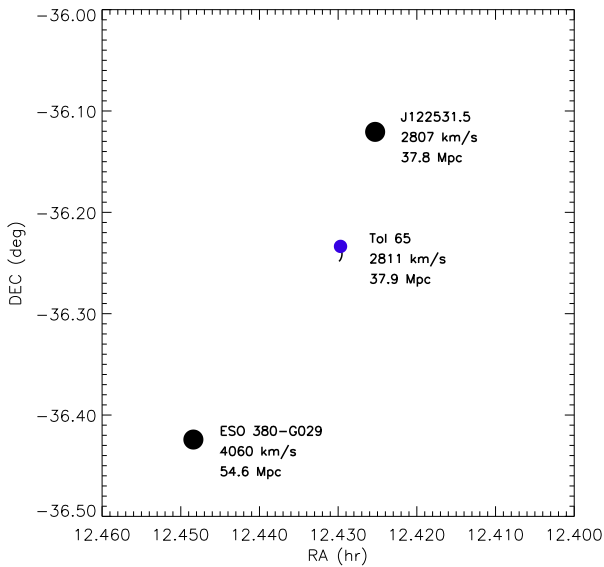}
  \caption{Field on the plane of the sky of Tol 65 (blue dot). We indicate in black the galaxies ESO 380-G029 (SB(s)m) and 
  GALEXMSC J122531.50-360715.3 (Irr). In this figure we show for each galaxy the heliocentric velocity 
  and the distance (3K CMB) obtained from NED.}
\label{figure_field_Tol65}
  \end{figure}

\subsubsection{Tidal interactions, mergers and/or gas infall}

In many respects, the tidal hypothesis is an attractive one for the origin of the tail in Tol 65,
given that tidal interactions enhance lopsidedness in low mass galaxies \citep{YozinBekki2014}.
However, H\,{\sc ii}/BCDs tend to populate low-density environments and are not associated with massive galaxies 
\citep[e.g.][]{TellesTerlevich1995,TellesMaddox2000,Noeske2001} suggesting that tidal interactions with massive 
companions could not be the dominant starburst triggering mechanism. 
Thus, the  cometary or tadpole shapes among these may not generally be major mergers.
We used the NASA/IPAC Extragalactic Database (NED) to search for nearby objects with measured systemic velocities. 
Figure \ref{figure_field_Tol65} shows the field on the plane of the sky of the galaxy. 
We find two galaxies, with measured velocities, in the field of Tol 65. The first one J122531.50-360715.3
is an irregular (dIrr) and likely dwarf galaxy with a difference between their respective 
systemic velocities of $\Delta$V = -4 km s$^{-1}$. 
ESO Digitized Sky Survey (DSS) images show that this object is highly disturbed with an apparent size
of $\sim$30$\arcsec$ equivalent with $\sim$6.2 kpc, which is two times the apparent size of Tol 65. 
The projected distance between these galaxies is r$_{p}\sim$100 kpc.
The other object in the field is the spiral galaxy ESO 380-G029. 
For this galaxy we found a $\Delta$V = 1249 km s$^{-1}$, thus it is unlikely that this galaxy 
is producing the tidally-disturbed tail present in the galaxy, because companions with 
$\Delta$V$>$ 500 km s$^{-1}$  have not a significant dynamical influence \citep{Noeske2001}.
It is interesting to note that the tail of the galaxy is oriented, in the plane of the sky, to this spiral galaxy. 
Possibly ESO 380-G029 had an stronger influence over Tol 65 in the past (several Gyr ago), 
thus the tail in the latter could be a remnant of that tidal interaction.
However, there is no observational evidence showing that this galaxy had an interaction 
in the past with Tol 65 to produce a tidal tail. 
\cite{Papaderos1999} using VLT images suggested the presence of a low-surface-brightness 
companion galaxy of Tol 65, but it was discarded by \cite{Izotov2004}. They found this object 
is in fact a background galaxy. Thus, if this extended structure has a tidal origin with a close companion(s)
it may be related with a faint and massive one that have not been identified by optical surveys. 
In any case, J122531.50-360715.3 remains as the only potential perturber. 
In this sense, the slight enhancement of the SFR in region no. 1 with respect to region no. 2 and the 
lopsidedness of those star-forming regions could be the consequence of this tidal interaction.
Although it is unlikely that another dwarf galaxy of low mass, as it seems, could have a tidal effect at this distance. 
From the present observations we cannot say much about the impact of J122531.50-360715.3 
over the physical properties of Tol 65 but we speculate based on the aforementioned 
arguments that the co-evolution on these systems play an important role in the enhancement 
of star-formation \citep[e.g.][]{Stierwalt2014} and evolution in many of these galaxies \citep{Papaderos2012}. 

Several studies have shown that a number of star-forming dwarf galaxies present filamentary H\,{\sc i} 
and/or extended structures, which may indicate a recent cold gas infall/accretion 
\cite[e.g.][]{Ekta2008,EktaChengalur2010a,Lelli2014} and/or minor merger/interaction 
\cite[e.g.][]{MartinezDelgado2012,Chengalur2015}. 
In this sense, the extended structure of this galaxy could be a remnant of one of these processes.
\cite{Verbeke2014} argue that if the stars are being formed as a consequence of the infall  
they will be formed along the path of the accretion. But, we did not observe the presence of star clusters
in the tail of the galaxy. Then, alternatively, the effect of these interactions has carried away in the recent past
an important fraction of gas and stars to form the stellar tail. 
The fact that $\sigma_{HII,body}\simeq\sigma_{HII,tail}\simeq\sigma_{HI}$ (see Sect. \ref{sect_velocity_field}) 
implies that the ionized gas still, on a global scale, retains the kinematic memory of its parental cloud 
and likely a common origin. While extended regions of diffuse gas are highly disturbed, probably due to unresolved 
expanding shells and the effects of massive star evolution.
Therefore, we interpreted the observed properties of the ISM and the morphology of the galaxy 
as due to a late-stage minor merger and/or inflow of metal-poor gas in the recent past of the galaxy. 

\subsection{On the metallicity of the ISM}\label{sect_disc_properties}

If the current star-forming episode in XMPs is the consequence of the infall of metal-poor gas, 
the less enriched gas dilutes the preexisting nuclear gas to produce a lower metallicity than would be obtained 
prior to the accretion. \cite{Filho2015} argue that such accretion flows could explain all the major XMP properties 
such as, i.e., isolation, lack of interaction/merger signatures, metal-poor H\,{\sc i} gas, metallicity inhomogeneities, etc.
In this sense, \cite{Sanchez2014a} argue that the variation of oxygen abundance along the major axis 
of two XMP star-forming dwarf galaxies can be interpreted as an early stage of assembling in disk galaxies 
with the star-formation sustained by external metal-poor gas accretion likely from the cosmic web \citep[][]{Sanchez2014b}. 
In \cite{Lagos2014} we studied one of the objects analyzed by \cite{Sanchez2014a}, 
the XMP galaxy HS 2236+1344. In this study, we found indications of variation of 
O and N/O associated with the less luminous star-formation region of that galaxy. 
But given the uncertainties related to those measurements we considered the abundances across the galaxy 
as fairly uniform. 
Since most of our H\,{\sc ii}/BCD galaxies \citep[][]{Lagos2009,Lagos2012,Lagos2014}, 
and XMP/BCDs from the literature \citep[][see references therein]{LagosPapaderos2013} studied so far 
are chemically homogeneous it seems that these galaxies have mixed up the metals with the pre-existing ISM 
before the current starburst appears. The newly synthesized metals from the current star-formation
episode reside in a hot gas phase (T $\gtrsim$ 10$^6$ K) and those will be dispersed probably in the
whole galaxy by the expansion of starburst-driven outflows. 
The energy injected by stellar winds and SNe may be able to eject part of the enriched gas into 
the intergalactic medium, but in most cases the expanding velocity  \citep[e.g.][and references therein]{vanEymeren2010} 
are not sufficient to allow the gas to escape from the potential well of the galaxies. 
In addition, inflows or accretion of relatively low metallicity gas could possibly dilute more easily 
the abundance of the gas on large scales. 
Assuming a sound speed of $\sim$12 km s$^{-1}$ (see Sect. \ref{sect_velocity_field}) as the expected speed 
at which mixing occurs in Tol 65, we found that over 1.5-2.0 kpc it yields $\sim$10$^8$yr, 
which is similar to the timescale required for cooling and dispersal of metals produced by massive stars 
\citep[e.g.][]{TenorioTagle1996} as observed in compact H\,{\sc ii}/BCD galaxies \cite[e.g. UM 408;][]{Lagos2009}.
Consequently, the predominant mechanism for metal transport may not be resolved in star-forming dwarf galaxies.
However, if we interpreted the observed properties of the ISM as due to a late stage-merger 
and/or inflow of metal-poor gas, the accreted gas must be almost fully diluted as seen in Fig. \ref{figure_SFR_OH}.
Therefore, the final mixture of gas thus depends on the relative enrichment of the acquired and pre-enriched gas.

\cite{Lebouteiller2009} showed that the neutral gas metallicity, in a sample of BCDs, 
is equal or lower than the ones found in the H\,{\sc ii} region.
This agree with the results obtained by \cite{Lebouteiller2013}, using the Cosmic Origin Spectrograph (COS) onboard HST,
in the sense that the H\,{\sc i} gas abundances is factor $\sim$2 lower than the warm gas abundance 
in \,{\sc i} Zw18 indicating that infall of, non fully pristine, metal-poor gas could be the responsible for its low metallicity.
Therefore, we cannot exclude that the tail in this XMP galaxy could be made from a gas with much lower metallicity 
than that in the warm gas phase. However, we still lack an adequate explanation about the evolutionary status of Tol 65 
since no spatially resolved H\,{\sc i} observations are available, but the infall/accretion of cold gas 
from the outskirts of the galaxy and/or minor merger/interaction with an small companion recently 
in the past of the galaxy could explain the main properties of the galaxy.
As mentioned above, the tail and the \textit{cometary morphology} 
could be the remnant of these interactions in which an appreciable fraction of gas has been 
carried away as the result of the tidal torques, then producing the mixing of metals through 
the ISM at large scales. In this sense, the scatter in Fig. \ref{figure_SFR_OH} indicates that the metals in the ISM were  
diluted in a relatively small timescale.
If the past infall of metal poor gas do not explain the low metallicity
of Tol 65, alternatively, it could be merely a consequence of the low star-formation efficiency 
due by the cosmic downsizing.

\section{Summary and Conclusions}\label{conclusions}

In this work we have presented VIMOS-IFU spectroscopy of the XMP H\,{\sc ii}/BCD galaxy
Tol 65. We studied the spatial distribution of properties (i.e., emission lines, abundances, kinematics, etc) 
through the ISM of the galaxy in an extended area of 13\arcsec$\times$13$\arcsec$ encompassing 
the two GH\,{\sc ii}Rs and most of the extended stellar tail.
Below, we summarize our results and conclusions:

\begin{enumerate}
       
\item This galaxy shows a clear cometary shape with a bright main body and an extended
      and diffuse stellar tail. We found that the current star-formation activity in Tol 65 
      started recently about 3-5 Myr ago. Our observations show the presence of an extended H$\alpha$ 
      emission in the tail of the galaxy. The mass of the ionized gas in the tail corresponds to
      $\sim$24 per cent of the total mass of the ionized gas in the galaxy.\\

\item The integrated properties in the main body of the galaxy, it to say, the extinction coefficient c(H$\beta$)=0.29, 
      oxygen abundance 12+log(O/H)=7.56$\pm$0.06, velocity dispersion $\sigma$=20.8$\pm$0.6 kms$^{-1}$ 
      obtained in this study agree, within the uncertainties, with the ones determined in the literature 
      \citep[e.g.][]{KunthSargent1983,Izotov2004,Guseva2011,BordaloTelles2011}.\\

\item The velocity field V$_r$ in Tol 65 is highly disturbed and it shows no global rotation pattern.
      We found $\sigma_{HII,body}\simeq \sigma_{HII,tail} \simeq \sigma_{HI}$ suggesting  that the gas
      still retains the kinematic memory of its parental cloud and a common origin. 
      While extended regions of diffuse gas are highly disturbed, probably due to unresolved expanding shells and the
      effects of massive star evolution.
      
\item  We did not observe a spatial variation of metal (O, N, Ne, Ar and S) in the ISM of the galaxy.
       This agrees with previous observations in other H\,{\sc ii} galaxies in the literature 
       \citep[][and references therein]{LagosPapaderos2013}.
       Although the evidence is far from conclusive, we favour the idea that the most likely mechanisms 
       to produce the flat abundance gradient and the cometary morphology in this XMP H\,{\sc ii}/BCD 
       is the infall/accretion of cold gas from the outskirts of the galaxy or minor merger/interaction 
       with an small companion recently in the past of the galaxy.
       This is in agreement with the scatter found between the gas metallicity (12+log(O/H)) and 
       star-formation rate at spaxel scales, in Fig. \ref{figure_SFR_OH}, which clearly indicates 
       that the metals in the ISM are almost fully diluted.\\

\end{enumerate}

Our VIMOS-IFU observation of Tol 65 provides a picture in which a late-stage minor merger and/or infall of gas 
explain most of the observed properties over the ISM. Therefore, we suggest that these global effects might be 
attributed as the main mechanism diminishing the preexisting metal of the galaxy \citep[e.g.][]{Lebouteiller2013}, 
then keeping the oxygen abundance (and other $\alpha$ elements) constant through the ISM at large scales. 
Clearly, H\,{\sc i} observations of Tol 65  with high spatial resolution and HST COS observations 
are needed in order to study the cool gas component and check whether the proposed mechanism are capable to produce 
the homogeneity of the ISM, the difference between the neutral and warm gas metallicity and the morphology 
of this XMP galaxy.

\section*{Acknowledgments}
We would like to thank the referee, Dr. Daniel Kunth, for his very valuable comments and suggestions.
This work was supported by Funda\c{c}\~ao para a Ci\^encia e a Tecnologia (FCT) through the
research grant UID/FIS/04434/2013. P.L. is supported by a Post-Doctoral grant SFRH/BPD/72308/2010, funded by FCT (Portugal).
R.D. gratefully acknowledges the support provided by the BASAL Center for Astrophysics and Associated 
Technologies (CATA), and by FONDECYT grant N. 1130528.
P.L., P.P., A.H., N.R. and J.M.G. acknowledge support by the FCT under project 
FCOMP-01-0124-FEDER-029170 (Reference FCT PTDC/FIS-AST/3214/2012), funded by the FEDER program. 
P.P. is supported by FCT through the Investigador FCT Contract No.
IF/01220/2013 and POPH/FSE (EC) by FEDER funding through the program COMPETE.
J.M.G. is supported by a Postdoctoral grant SFRH/BPD/66958/2009, funded by FCT (Portugal).
We acknowledge support by the exchange programme ‘Study of Emission-Line Galaxies with Integral-
Field Spectroscopy’ (SELGIFS, FP7-PEOPLE-2013-IRSES-612701), funded by the EU through the
IRSES scheme.
This research has made use of the NASA/IPAC Extragalactic Database (NED) which is operated 
by the Jet Propulsion Laboratory, California Institute of Technology, under contract with 
the National Aeronautics and Space Administration.



\begin{thebibliography}{99}

\bibitem[\protect\citeauthoryear{Atek et al.}{2008}]{Atek2008} Atek, H., Kunth, D., Hayes, M. \"{O}stlin, G., 
Mas-Hesse, J. M. 2008, A\&A, 488, 491

\bibitem[\protect\citeauthoryear{Baldwin, Phillips \& Terlevich}{1981}]{Baldwin1981} Baldwin, J. A., Phillips, M. M., 
Terlevich  R. 1981, PASP, 93, 5   

\bibitem[\protect\citeauthoryear{Bekki}{2008}]{Bekki2008} Bekki, K. 2008, MNRAS, 388, 10

\bibitem[\protect\citeauthoryear{Bordalo \& Telles}{2011}]{BordaloTelles2011} Bordalo, V., Telles, E. 2011, ApJ, 735, 52

\bibitem[\protect\citeauthoryear{Bordalo, Plana \& Telles}{2009}]{Bordalo2009} Bordalo, V., Plana, H., Telles E. 2009, 
ApJ, 696, 1668

\bibitem[\protect\citeauthoryear{Bournaud \& Elmegreen}{2009}]{Bournaud2009} Bournaud, F., Elmegreen, B. G. 2009, ApJ, 
694, 158

\bibitem[\protect\citeauthoryear{Cardelli, Clayton \& Mathis}{1989}]{Cardelli1989} Cardelli, J. A., Clayton, G. C., Mathis, J. S. 
1989, ApJ, 345, 245 

\bibitem[\protect\citeauthoryear{Calzetti et al.}{2007}]{Calzetti2007} Calzetti, D. et al. 2007, ApJ, 666, 870

\bibitem[\protect\citeauthoryear{Cervi\~no \& Mas-Hesse}{1994}]{CervinoMasHesse1994} Cervi\~no, M., Mas-Hesse, J. M. 1994,
A\&A, 284, 749

\bibitem[\protect\citeauthoryear{Copetti, Pastoriza \& Dottori}{1986}]{Copetti1986} Copetti, M. V. F., Pastoriza, M. G., 
Dottori, H. A. 1986, A\&A, 156, 111

\bibitem[\protect\citeauthoryear{Charlot \& Fall}{1993}]{CharlotFall1993} Charlot, S., Fall, S. M. 1993, ApJ, 415, 580

\bibitem[\protect\citeauthoryear{Chengalur et al.}{2015}]{Chengalur2015} Chengalur, J. N., Pustilnik, S. A., Makarov, D. I.,
Perepelitsyna, Y. A., Safonova, E. S., Karachentsev, I. D. 2015, MNRAS, 448, 1634

\bibitem[\protect\citeauthoryear{Dekel \& Birnboim}{2006}]{DekelBirnboim2006} Dekel, A., Birnboim, Y. 2006, MNRAS, 368, 2

\bibitem[\protect\citeauthoryear{Dekel et al.}{2009}]{Dekel2009} Dekel, A., Birnboim, Y., Engel, G., Freundlich, J.,
Goerdt, T., Mumcuoglu, M., et al. 2009, Nature, 457, 451

\bibitem[\protect\citeauthoryear{De Robertis, Dufour \& Hunt}{1987}]{DeRobertis1987} De Robertis, M. M., Dufour, R. J., 
Hunt, R. W. 1987, JRASC, 81, 195

\bibitem[\protect\citeauthoryear{Dottori}{1981}]{Dottori1981} Dottori, H. A. 1981, Ap\&SS, 80, 267

\bibitem[\protect\citeauthoryear{Duc, Bournaud \& Massetet al.}{2004}]{Duc2004} Duc, P.-A., Bournaud, F., Masset, F. 2004, A\&A, 
427, 803

\bibitem[\protect\citeauthoryear{Duc et al.}{2011}]{Duc2011} Duc, P.-A., Cuillandre, J.-C., Serra, P., et al. 2011, 
MNRAS, 417, 863

\bibitem[\protect\citeauthoryear{Ekta \& Chengalur}{2010a}]{EktaChengalur2010a} Ekta, B., Chengalur, J. N. 2010a, MNRAS, 
403, 295

\bibitem[\protect\citeauthoryear{Ekta \& Chengalur}{2010b}]{EktaChengalur2010b} Ekta, B., Chengalur, J. N. 2010b, MNRAS, 
406, 1238

\bibitem[\protect\citeauthoryear{Ekta, Chengalur \& Pustilnik}{2008}]{Ekta2008} Ekta, B., Chengalur, J. N., Pustilnik, S. A. 
2008, MNRAS, 391, 881

\bibitem[\protect\citeauthoryear{Elmegreen \& Elmegreen}{2010}]{ElmegreenElmegreen2010} Elmegreen, B. G., Elmegreen, D. M. 2010
ApJ, 722,1895

\bibitem[\protect\citeauthoryear{Elmegreen et al.}{2012}]{Elmegreen2012} Elmegreen, D. M., Elmegreen, B. G.,
S\'anchez Almeida, J., Mu\~noz-Tu\~n\'on, C., Putko, J., Dewberry, J. 2012, ApJ, 750, 95

\bibitem[\protect\citeauthoryear{Filho et al.}{2013}]{Filho2013} Filho, M. E., Winkel, B., S\'anchez Almeida, J., 
Aguerri, J. A., Amor\'{\i}n, R., Ascasibar, Y., Elmegreen, B. G., Elmegreen, D. M., Gomes, J. M., Humphrey, A., 
Lagos, P., Morales-Luis, A. B., Mu\~noz-Tu\~n\'on, C., Papaderos, P., V\'{\i}lchez, J. M. 2013, A\&A, 558, 18

\bibitem[\protect\citeauthoryear{Filho et al.}{2015}]{Filho2015} Filho, M. E., S\'anchez Almeida, J., Mu\~noz-Tu\~n\'on, C.,
Nuza, S. E., Kitaura, F., Hess, S. 2015, ApJ, 802, 82

\bibitem[\protect\citeauthoryear{Giavalisco, Koratkar \& Calzetti}{1996}]{Giavalisco1996} Giavalisco, M., Koratkar, A.,
Calzetti, D. 1996, ApJ, 466, 831

\bibitem[\protect\citeauthoryear{Gil de Paz, Madore \& Pevunova}{2003}]{GildePaz2003} Gil de Paz, A., Madore, B. F., Pevunova, O. 
2003, ApJS, 147, 29

\bibitem[\protect\citeauthoryear{Guaita et al.}{2015}]{Guaita2015} Guaita, L., Melinder, J., Hayes, M., 
et al.\ 2015, ApJ, 576, A51 

\bibitem[\protect\citeauthoryear{Guseva et al.}{2011}]{Guseva2011} Guseva, N. G., Izotov, Y. I., Stasinska, G., Fricke, K. J.,
Henkel, C., Papaderos, P. 2011, A\&A 529, 149 

\bibitem[\protect\citeauthoryear{Guseva et al.}{2012}]{Guseva2012} Guseva, N. G., Izotov, Y. I., Fricke, K. J.,
Henkel, C. 2012, A\&A, 541, 115

\bibitem[\protect\citeauthoryear{Humphrey et al.}{2013}]{Humphrey2013} Humphrey, A., Vernet J., 
Villar-Mart{\'{\i}}n, M., di Serego Alighieri, S., Fosbury, R.~A.~E., Cimatti, A., 2013, ApJ, 768, L 

\bibitem[\protect\citeauthoryear{Izotov et al.}{1997}]{Izotov1997} Izotov, Y. I., Lipovetsky, V. A., Chaffee, F.H., 
Foltz, C. B., Guseva, N. G., Kniazev, A. Y. 1997, ApJ, 476, 698

\bibitem[\protect\citeauthoryear{Izotov et al.}{2001}]{Izotov2001} Izotov, Y. I., Papaderos, P., Guseva, N. G., Fricke, K. J.,
Thuan, T. X. 2001, Izotov,Y. I., Chaffee, F. H., Green, R. F. 2001, ApJ, 562, 727

\bibitem[\protect\citeauthoryear{Izotov et al.}{2004}]{Izotov2004} Izotov, Y. I., Papaderos, P., Guseva, N. G., 
Fricke, K. J., Thuan, T. X. 2004, A\&A, 421, 539

\bibitem[\protect\citeauthoryear{Izotov et al.}{2006}]{Izotov2006} Izotov, Y. I., Schaerer, D., Blecha, A., Royer, F.,
Guseva, N. G., North, P. 2006, A\&A, 459, 71

\bibitem[\protect\citeauthoryear{Kennicutt}{1998}]{Kennicutt1998} Kennicutt, R. C. 1998 ARA\&A, 36, 189

\bibitem[\protect\citeauthoryear{Kingsburgh \& Barlow}{1994}]{KingsburhBarlow1994} Kingsburgh, R. L., Barlow, M. J. 1994, 
MNRAS, 271, 257

\bibitem[\protect\citeauthoryear{Kobulnicky \& Skillman}{1997}]{KobulnickySkillman1997} Kobulnicky, H. A., Skillman, E. D. 1997, 
ApJ, 489, 636

\bibitem[\protect\citeauthoryear{Kobulnicky \& Skillman}{1998}]{KobulnickySkillman1998} Kobulnicky, H. A., Skillman, E. D. 
1998, ApJ, 497, 601

\bibitem[\protect\citeauthoryear{Kunth \& Sargent}{1983}]{KunthSargent1983} Kunth, D., Sargent, W. L. W. 1983, ApJ, 273, 81

\bibitem[\protect\citeauthoryear{Kunth \& Jourbert}{1985}]{KunthJourbert1985} Kunth, D., Joubert, M. 1985, A\&A,142, 411

\bibitem[\protect\citeauthoryear{Kunth et al.}{1994}]{Kunth1994} Kunth, D., Lequeux, J., Sargent, W. L. W., Viallefond, F. 
1994, A\&A, 282, 709

\bibitem[\protect\citeauthoryear{Kunth et al.}{1998}]{Kunth1998}Kunth, D., Mas-Hesse, J. M., Terlevich, E., Terlevich, R.,
Lequeux, J., Fall, S. M. 1998, A\&A, 334, 11

\bibitem[\protect\citeauthoryear{Kunth \& \"{O}stlin}{2000}]{KunthOstlin2000} Kunth, D., \"{O}stlin, G. 2000, ARA\&A, 10, 1

\bibitem[\protect\citeauthoryear{Lagerholm et al.}{2012}]{Lagerholm2012} Lagerholm, C., Kuntschner, H., Cappellari, M., 
Krajnovic, D., McDermid, R. M., Rejkuba, M. 2012, A\&A, 541, 82

\bibitem[\protect\citeauthoryear{Lagos et al.}{2007}]{Lagos2007} Lagos, P., Telles, E., Melnick, J. 2007, A\&A, 476, 89

\bibitem[\protect\citeauthoryear{Lagos et al.}{2009}]{Lagos2009} Lagos, P., Telles, E., Mu\~noz-Tu\~n\'on, C., Carrasco, E. R., 
Cuisinier, F., Tenorio-Tagle, G. 2009, AJ, 137, 5068

\bibitem[\protect\citeauthoryear{Lagos et al.}{2011}]{Lagos2011} Lagos, P., Telles, E., Nigoche-Netro, A., Carrasco, E. R.  2011, 
AJ, 142, 162

\bibitem[\protect\citeauthoryear{Lagos et al.}{2012}]{Lagos2012} Lagos, P., Telles, E., Nigoche-Netro A., Carrasco, E. R. 2012, 
MNRAS, 427, 740

\bibitem[\protect\citeauthoryear{Lagos \& Papaderos}{2013}]{LagosPapaderos2013} Lagos, P., Papaderos, P. 2013, AdAst, 2013, 20

\bibitem[\protect\citeauthoryear{Lagos et al.}{2014}]{Lagos2014} Lagos, P., Papaderos, P., Gomes, J. M., Smith, A. V., 
Vega, L. R. 2014, A\&A, 569, 110

\bibitem[\protect\citeauthoryear{Lebouteiller et al.}{2009}]{Lebouteiller2009} Lebouteiller, V., Heap, S., Hubeny, I., Kunth, D.
2009, A\&A, 494, 915

\bibitem[\protect\citeauthoryear{Lebouteiller et al.}{2013}]{Lebouteiller2013} Lebouteiller, V., Heap, S., Hubeny, I., Kunth, D.
2013, A\&A, 553, 16

\bibitem[\protect\citeauthoryear{Lequeux et al.}{1995}]{Lequeux1995} Lequeux, J., Kunth, D., Mas-Hesse, J. M., Sargent, W. L. W.
1995, A\&A, 301, 18

\bibitem[\protect\citeauthoryear{Lee \& Skillman}{2004}]{LeeSkillman2004} Lee, H., Skillman, E. D. 2004, ApJ, 614, 698

\bibitem[\protect\citeauthoryear{Leitherer et al.}{1999}]{Leitherer1999} Leitherer, C., et al. 1999, ApJS, 123, 3

\bibitem[\protect\citeauthoryear{Lelli, Verheijen \& Fraternali}{2014}]{Lelli2014} Lelli, F., Verheijen, M., Fraternali, F. 
2014, MNRAS, 445, 1694

\bibitem[\protect\citeauthoryear{Loose \& Thuan}{1986}]{LooseThuan1986} Loose, H.-H., Thuan, T. X. 1986, ApJ, 309, 59

\bibitem[\protect\citeauthoryear{L\'opez-S\'anchez}{2010}]{LopezSanchez2010} L\'opez-S\'anchez, \'A. R. 2010, A\&A, 521, 63

\bibitem[\protect\citeauthoryear{Mart\'{\i}nez-Delgado et al.}{2012}]{MartinezDelgado2012} Mart\'{\i}nez-Delgado, D., 
Romanowsky, A. J., Gabany, R. J., et al. 2012, ApJ, 748, 24

\bibitem[\protect\citeauthoryear{McQuinn et al.}{2009}]{McQuinn2009} McQuinn, K. B. W., Skillman, E. D., Cannon, J. M.,
Dalcanton, J. J., Dolphin, A., Stark, D., Weisz, D. 2009, ApJ, 695, 561

\bibitem[\protect\citeauthoryear{Moiseev, Tikhonov \& Klypin}{2015}]{Moiseev2015} Moiseev, A. V., Tikhonov, A. V., Klypin, A. 
2015, MNRAS, 449, 3568

\bibitem[\protect\citeauthoryear{Noeske et al.}{2001}]{Noeske2001} Noeske, K. G., Iglesias-P\'aramo, J., Vilchez, J. M., 
Papaderos, P., Fricke, K. J. 2001, A\&A, 371, 806

\bibitem[\protect\citeauthoryear{Noeske et al.}{2003}]{Noeske2003} Noeske, K. G., Papaderos, P., Cair\'os, L. M., Fricke, K. J.
2003, A\&A, 410, 481

\bibitem[\protect\citeauthoryear{Osterbrock \& Ferland}{2006}]{OsterbrockFerland2006} Osterbrock, D. E., Ferland, G. J. 2006, 
Astrophysics of gaseous nebulae and active galactic nuclei, 2nd. edn. (Sausalito, CA: University Science Books)


\bibitem[\protect\citeauthoryear{Papaderos et al.}{1999}]{Papaderos1999} Papaderos, P., Fricke, K. J., Thuan, T. X.,
Izotov, Y. I., Nicklas, H. 1999, A\&A, 352, 57

\bibitem[\protect\citeauthoryear{Papaderos et al.}{2008}]{Papaderos2008} Papaderos, P., Guseva, N. G., Izotov, Y. I., 
Fricke, K. J. 2008, A\&A, 491, 113

\bibitem[\protect\citeauthoryear{Papaderos \& \"{O}stlin}{2012}]{PapaderosOstlin2012} Papaderos, P., \"{O}stlin, G. 2012, 
A\&A, 537, A126

\bibitem[\protect\citeauthoryear{Papaderos}{2012}]{Papaderos2012} Papaderos, P. 2012, in Dwarf Galaxies: Keys to Galaxy 
Formation and Evolution, proceedings of JENAM\ 2010 (Lisbon, September 2010), P. Papaderos, S. Recchi, G. Hensler (eds.), 
Springer Verlag, p. 321 

\bibitem[\protect\citeauthoryear{Papaderos et al.}{2013}]{Papaderos2013} Papaderos, P., Gomes, J. M., V\'{\i}lchez, J. M. 
et al. 2013, A\&A Letters, 555, 1

\bibitem[\protect\citeauthoryear{Pagel et al.}{1992}]{Pagel1992} Pagel, B. E. J., Simonson, E. A., Terlevich, R. J., 
Edmunds, M. G. 1992, MNRAS, 255, 325

\bibitem[\protect\citeauthoryear{Partridge \& Peebles}{1967}]{PartridgePeebles1967} Partridge, R. B., Peebles, P. J. E. 1967, 
ApJ, 147, 868

\bibitem[\protect\citeauthoryear{P\'erez-Montero et al.}{2011}]{PerezMontero2011} P\'erez-Montero, E., V\'{\i}lchez, J. M., 
Cedr\'es, B., H$\ddot{a}$gele, G. F., Moll\'a, M., Kehrig, C., D\'{\i}az, A. I., Garc\'{\i}a-Benito, R., 
Mart\'{\i}n-Gord\'on, D. 2011, A\&A, 532, 141

\bibitem[\protect\citeauthoryear{Pustilnik \& Martin}{2007}]{PustilnikMartin2007} Pustilnik, S. A., Martin, J.-M. 2007, A\&A, 
464, 859

\bibitem[\protect\citeauthoryear{Pustilnik et al.}{2001}]{Pustilnik2001} Pustilnik, S. A., Kniazev, A. Y., Lipovetsky, V. A., 
Ugryumov, A. V. 2001, A\&A, 373, 24

\bibitem[\protect\citeauthoryear{Rich et al.}{2012}]{Rich2012} Rich, J. A., Torrey, P., Kewley, L. J.,
Dopita, M. A., Rupke, D. S. N.  2012, ApJ, 753, 5 

\bibitem[\protect\citeauthoryear{Roy \& Kunth}{1995}]{RoyKunth1995} Roy, J.-R., Kunth, D. 1995, A\&A, 294, 432

\bibitem[\protect\citeauthoryear{Rupke, Kewley \& Barnes}{2010}]{Rupke2010} Rupke, D. S. N., Kewley, L. J., Barnes, J. E. 2010, 
ApJ, 710, 156

\bibitem[\protect\citeauthoryear{Shaw \& Dufour}{1994}]{ShawDufour1994} Shaw, R. A., Dufour, R. J. 1994, ASPC, 61, 327

\bibitem[\protect\citeauthoryear{Sanders et al.}{1988}]{Sanders1988} Sanders, D. B., Soifer, B. T., Elias, J. H., et al. 1988,
ApJ, 325, 74

\bibitem[\protect\citeauthoryear{S\'anchez Almeida et al.}{2013}]{Sanchez2013} S\'anchez Almeida, J., Mu\~noz-Tu\'n\'on, C., 
Elmegreen, D. M., Elmegreen, B. G., M\'endez-Abreu, J. 2013, ApJ, 767, 74

\bibitem[\protect\citeauthoryear{S\'anchez Almeida et al.}{2014a}]{Sanchez2014a} S\'anchez Almeida, J., Morales-Luis, A. B., 
Mu\~noz-Tun\'on, C., Elmegreen, D. M., Elmegreen, B. G., M\'endez-Abreu, J. 2014a, ApJ, 783, 45

\bibitem[\protect\citeauthoryear{S\'anchez Almeida et al.}{2014b}]{Sanchez2014b} S\'anchez Almeida, J., Elmegreen, B. G.,
Mu\~noz-Tu\~n\'on, C., Elmegreen, D. M. 2014b, A\&ARv, 22, 71

\bibitem[\protect\citeauthoryear{Smoker et al.}{2000}]{Smoker2000} Smoker, J. V., Davies, R. D., Axon, D. J., Hummel, E. 2000, 
A\&A, 361, 19

\bibitem[\protect\citeauthoryear{Springel}{2000}]{Springel2000} Springel, V. 2000, MNRAS, 312, 859

\bibitem[\protect\citeauthoryear{Stierwalt et al.}{2014}]{Stierwalt2014} Stierwalt, S., Besla, G., Patton, D., Johnson, K.,
Kallivayalil, N., Putman, M., Privon, G., Ross, G. 2015, ApJ, 805, 2

\bibitem[\protect\citeauthoryear{Straughn et al.}{2006}]{Straughn2006} Straughn, A. N., Cohen, S. H., Ryan, R. E., et al. 
2006, ApJ, 639, 724

\bibitem[\protect\citeauthoryear{Taylor, Brinks \& Skillman}{1993}]{Taylor1993} Taylor, C., Brinks, E., Skillman, E. D.
1993, AJ, 105, 128

\bibitem[\protect\citeauthoryear{Telles \& Terlevich}{1995}]{TellesTerlevich1995} Telles, E., Terlevich, R. 1995, MNRAS, 275, 1

\bibitem[\protect\citeauthoryear{Telles \& Maddox}{2000}]{TellesMaddox2000} Telles, E., Maddox, S. 2000, MNRAS, 311, 307

\bibitem[\protect\citeauthoryear{Tenorio-Tagle}{1996}]{TenorioTagle1996} Tenorio-Tagle, G. 1996, AJ, 111, 1641

\bibitem[\protect\citeauthoryear{Tenorio-Tagle et al.}{1999}]{TenorioTagle1999} Tenorio-Tagle, G., Silich, S. A., Kunth, D., 
Terlevich, E., Terlevich, R. 1999, MNRAS, 309, 332

\bibitem[\protect\citeauthoryear{Thuan \& Izotov}{1997}]{ThuanIzotov1997} Thuan, T. X., Izotov, Y. I. 1997, ApJ, 489, 623

\bibitem[\protect\citeauthoryear{Thuan, Izotov\& Lipovetsky}{1997}]{Thuan1997} Thuan, T. X., Izotov, Y. I., Lipovetsky, V. A. 
1997, ApJ, 477, 661

\bibitem[\protect\citeauthoryear{Toomre}{1977}]{Toomre1977} Toomre, A. 1977, ARA\&A, 15, 437 

\bibitem[\protect\citeauthoryear{van Eymeren et al.}{2010}]{vanEymeren2010} van Eymeren, J., Koribalski, B, S.,
L\'opez-S\'anchez, \'A. R., Dettmar, R.-J., Bomans, D. J. 2010, MNRAS, 407, 113

\bibitem[\protect\citeauthoryear{Verbeke et al.}{2014}]{Verbeke2014} Verbeke, R., De Rijcke, S., Koleva, M., Cloet-Osselaer, A.,
Vandenbroucke, B., Schroyen, J. 2014, MNRAS, 442, 1830

\bibitem[\protect\citeauthoryear{Villar-Mart{\'{\i}}n et al.}{2007}]{VillarMartin2007} Villar-Mart{\'{\i}}n, M., Humphrey, A., 
De Breuck, C., Fosbury, R., Binette, L., Vernet, J. 2007, MNRAS, 375, 1299 

\bibitem[\protect\citeauthoryear{Walsh \& Roy}{1990}]{WalshRoy1990} Walsh, J. R., Roy, J. R. 1990, ESOC, 34, 95

\bibitem[\protect\citeauthoryear{Yozin \& Bekki}{2014}]{YozinBekki2014} Yozin, C., Bekki, K. 2014, MNRAS, 439, 1948

\end{thebibliography}
\end{document}